\documentclass[reprint,amsmath,amssymb,aps]{revtex4-1}
\usepackage{graphicx}
\usepackage{dcolumn}
\usepackage{bm}
\usepackage{natbib}
\usepackage[T2A,T1]{fontenc}
\newcommand{\Sha}{\mbox{\usefont{T2A}{\rmdefault}{m}{n}\CYRSH}}

\usepackage[colorlinks,linkcolor=black,anchorcolor=blue,citecolor=blue,urlcolor=blue]{hyperref}

\begin{document}
\title{Pervasive orientational and directional locking at geometrically heterogeneous sliding interfaces}
\author{Xin Cao$^{1}$}
\author{Emanuele Panizon$^{1}$}
\author{Andrea Vanossi$^{2,3}$}
\author{Nicola Manini$^{4}$}
\author{Erio Tosatti$^{2,3,5}$}
\author{Clemens Bechinger$^{1}$}
\email{clemens.bechinger@uni-konstanz.de}
\affiliation{$^1$Fachbereich Physik, Universit\"at Konstanz, 78464 Konstanz, Germany}
\affiliation{$^2$International School for Advanced Studies (SISSA), Via Bonomea 265, 34136 Trieste, Italy}
\affiliation{$^3$CNR-IOM Democritos National Simulation Center, Via Bonomea 265, 34136 Trieste, Italy}
\affiliation{$^4$Dipartimento di Fisica, Universit\`a degli Studi di Milano, Via Celoria 16, 20133 Milano, Italy}
\affiliation{$^5$International Centre for Theoretical Physics (ICTP), Strada Costiera 11, 34151 Trieste, Italy}
\date{\today}
\begin{abstract}
Understanding the drift motion and dynamical locking of crystalline clusters on patterned substrates is important for the diffusion and manipulation of nano- and micro-scale objects on surfaces. In a previous work, we studied the orientational and directional locking of colloidal two-dimensional clusters with triangular structure driven across a triangular substrate lattice. Here we show with experiments and simulations that such locking features arise for clusters with arbitrary lattice structure sliding across arbitrary regular substrates. Similar to triangular-triangular contacts, orientational and directional locking are strongly correlated via the real- and reciprocal-space moir\'e patterns of the contacting surfaces. Due to the different symmetries of the surfaces in contact, however the relation between the locking orientation and the locking direction becomes more complicated compared to interfaces composed of identical lattice symmetries. We provide a generalized formalism which describes the relation between the locking orientation and locking direction with arbitrary lattice symmetries.
\end{abstract}

\maketitle

\begin{figure*}
  \centering
  \includegraphics[width=2.0\columnwidth]{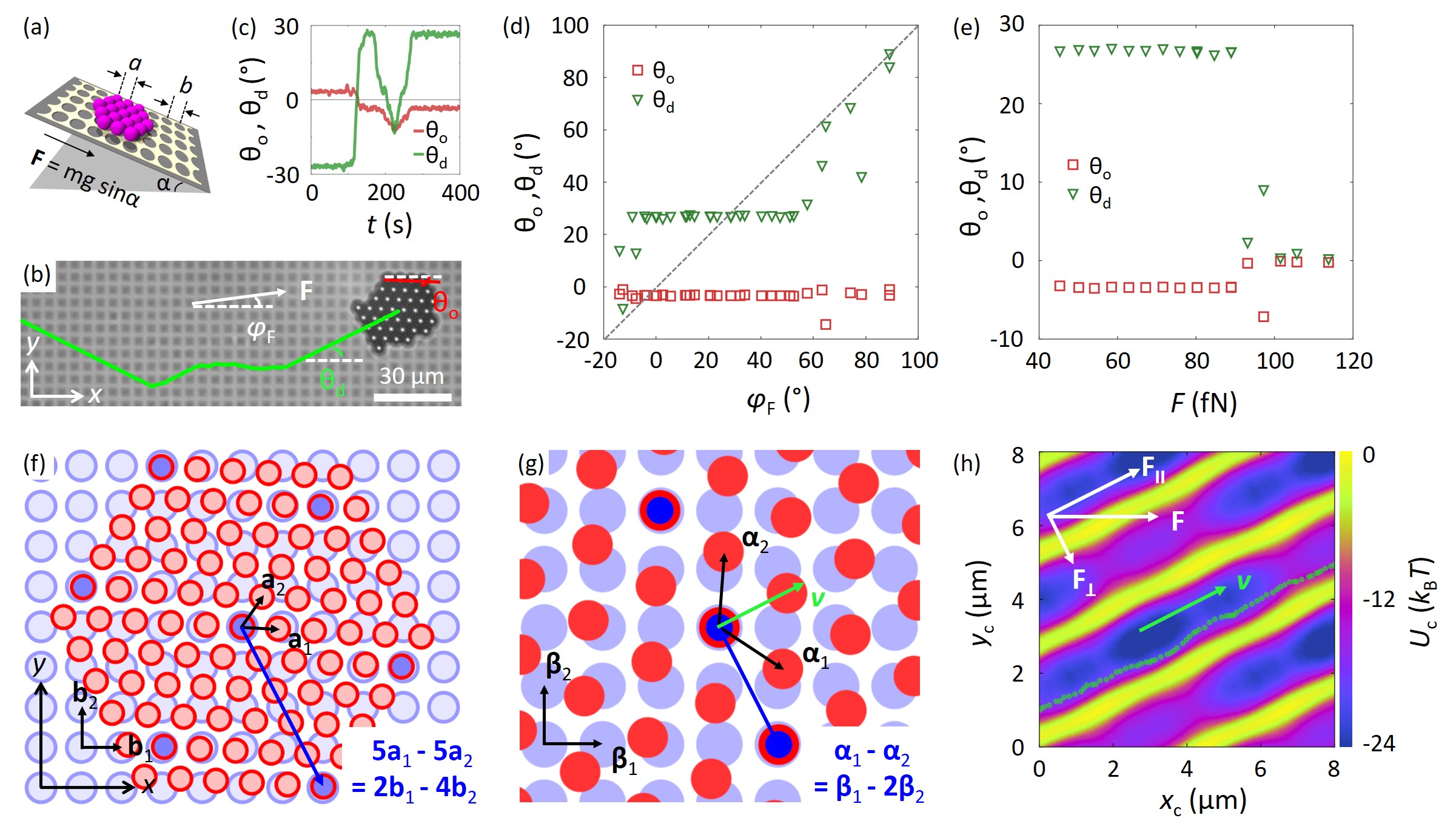}
\caption{\textbf{Experimental observation of orientational and directional locking of triangularly packed colloid clusters of spacing $a = 4.45 ~\mu\rm{m}$ on a square lattice with $b = 5.0 ~\mu\rm{m}$.} (a) The experimental setup. (b) Microscopy image of a colloidal cluster of $N$ = 40 particles. The green line is the cluster's center-of-mass trajectory under the driving force $F = 72$~fN and $\varphi_\textrm{F} = 0^\circ$. (c) The orientation of the cluster $\theta_\textrm{o}$ and its direction of motion $\theta_\textrm{d}$ as a function of time $t$ for the cluster in (b). (d) The measured $\theta_\textrm{o}$ and $\theta_\textrm{d}$ as a function of the force orientation $\varphi_\textrm{F}$ for 12 clusters of $N = 30$--120 particles. Each data point is evaluated from a cluster trajectory of $\sim 50~\mu\rm{m}$ length. Dashed line indicates $\theta_\textrm{d}$ = $\varphi_\textrm{F}$. The driving force amplitude $F$ is between 54~fN and 72~fN. (e) The measured $\theta_\textrm{o}$ and $\theta_\textrm{d}$ as a function of $F$ for 7 clusters of $N = 60$--140 particles in experiments, given $\varphi_\textrm{F}$ = 0$^{\circ}$. Data points are not measured at $F <  \sim 40$~fN because the clusters are not moving. (f) Illustration of the moir\'e pattern when the angle between $\textbf{a}_1$ and $\textbf{b}_1$ is the energetically optimal $\theta_\textrm{o}$ = $-3.4^{\circ}$. The blue arrow indicates the CLV $5\textbf{a}_1 - 5\textbf{a}_2 \approx 2\textbf{b}_1 - 4\textbf{b}_2$. (g) Illustration of the moir\'e pattern of the reciprocal lattices for the lattices in (f). The blue arrow indicates $\bm{\alpha}_1 - \bm{\alpha}_2 \approx \bm{\beta}_1 - 2\bm{\beta}_2$, which is perpendicular to the cluster's directionally-locked velocity $v$. (h) The calculated per-particle potential energy $U(\textbf{r}_\textrm{c}, \theta_\textrm{o}=-3.43^{\circ}) = (1/N) \sum_j V(\textbf{r}_j)$ acting on the cluster in (f), as a function of its center-of-mass position $\textbf{r}_\textrm{c} = (x_\textrm{c},  y_\textrm{c})$ on the square surface. The green dotted line is the center-of-mass trajectory of the cluster in (b,c) for a period of time from $t=312$ s to $t=335$ s, data points are taken at 0.33 s interval.}
\label{fig1}
\end{figure*}

\section{Introduction}
The dynamical behavior of crystalline adsorbates on ordered surfaces is relevant for a variety of condensed-matter-related problems ranging from nanofriction \cite{dienwiebel2004prl,filippov2008prl,guerra2010nmat,dietzel2013prl,song2018nmat,vanossi2020nc} to surface-based transport \cite{jensen1999rmp,lewis2000prb,ala2002ap,lebedeva2010prb} and manipulations \cite{lopinski2000nature,trillitzsch2018prb,kim2011nano}.
In contrast to a single molecule whose motion is, in general, simply related to the high symmetry axes of the underlying corrugated substrate \cite{reichhardt1999prl,balvin2009prl,stoop2020prl}, the rheology of a crystalline cluster on a solid surface can be much more complicated due to the frustrated cluster-surface interactions arising from the competing symmetries, commensurabilities and rotation-translation couplings.
For example, friction anisotropy is frequently observed in scanning friction-force microscopy experiments where nanoparticles or crystalline flakes are pushed across a crystalline surface \cite{sheehan1996science,lucas2009nmat,balakrishna2014prb}, and anomalous diffusion has been reported in surface-based diffusions of crystalline nanoparticles \cite{luedtke1999prl,metzler2000pr,maruyama2004prb}. 

Colloidal particles are 'magnified atoms' with easily accessible mesoscopic time and length scales. In addition, relevant interactions can be tuned easily in colloidal experiments, which distinguishes them as model systems to study e.g. phase transitions of crystals and glasses \cite{gasser2010cpc,wang2012science,hunter2012rpp}, diffusion \cite{wei2000science}, self-assembly \cite{rogers2016nrmat}, nanofriction \cite{bohlein2012nmat,brazda2018prx}, plastic deformation \cite{cao2020nc} and properties of frustrated systems \cite{ortiz2016nc,libal2018nc}. In a previous work \cite{cao2019np}, we investigated the motional behavior of triangular colloidal crystalline clusters driven across a corrugated triangular crystalline surface. Compared with nanoscale crystalline islands or flakes composed of atoms and molecules as building blocks, mesoscopic ordered clusters made of colloid particles can be manipulated easily with an external field.
Besides, their translational and rotational dynamics relative to the underneath surface could be monitored with a high precision scarcely accessible with atomic clusters. There we found that the clusters are often locked to an orientation determined by the potential energy minimum in the cluster/substrate angular misalignment. Accompanied by the orientational locking, we found that the direction of motion of the colloidal clusters becomes  locked (directional locking) to specific, generally non-trivial directions determined by the lattice mismatch and cluster size.
We obtained in addition a simple geometrical relation between the locking orientation and the locking direction, all in the special case where both the colloidal cluster and the periodic substrate have triangular symmetry.
However, it is not obvious whether orientational/directional locking can be observed in more general interfaces composed of ordered lattices with different geometry.

Here we show that orientational and directional locking can occur for any contact that involves periodic or quasi-periodic lattices.
We start with experiments for the relatively simple case of a triangular-lattice cluster on square-lattice substrate. 
Then, we generalize the results to arbitrary periodic lattices. We show that, regardless of the symmetry of the surfaces in contact, orientational and directional locking remain strongly correlated via the moir\'e patterns in real and in reciprocal space, respectively. 
As a result of the difference in the cluster/substrate lattice symmetries, however, the relation between the locking orientation and the locking direction of clusters becomes considerably more complex compared to the case of identical lattice symmetries. 
In addition to periodic lattices, we also show that crystalline clusters can develop directional locking on quasiperiodic lattices.

\section{Experiments}
To create colloidal crystalline clusters, the colloidal spheres (Dynabeads with diameters 4.45 $\mu\rm{m}$) are dispersed into a polyacrylamide (PAAm)-water solution (0.02\% PAAm in mass).
The PAAm molecules $\rm{(-CH_2CHCONH_2-)_n}$ are sufficiently large (molar mass 18,000,000 g/mol) that they can be physisorbed at the surfaces of two or more colloidal particles simultaneously.
This leads to so-called bridging flocculation \cite{weissenborn1994ijmp,mcguire2006jcis}. Such bridging-induced attraction tightly binds the colloidal particles together, thus forming rigid colloidal two-dimensional (2D) clusters with lattice spacing $a$ = 4.45 $\mu\rm{m}$.
Because of the random collision-and-capture growth mechanism \cite{cao2019np}, clusters of different shapes and sizes (up to $\sim 400$ particles) are formed. Substrates with periodically and quasiperiodically arranged cylindrical wells are fabricated by photolithography.
The wells have a diameter $\sim 3.8~\mu\rm{m}$ and depth $\sim 80$~nm and are arranged as a square lattice with different lattice spacings $b$ = 4.8, 5.0, 5.4, 6.2 $\mu\rm{m}$, or a quasiperiodic lattice composed of fat and slim rhombi.
The colloidal suspension is injected into a rectangular cell with 300 $\mu\rm{m}$ in height. The bottom plate of the cell contains the 20 mm $\times$ 30 mm corrugated surface. As illustrated in Fig.~1(a), due to gravity, the colloidal clusters sediment on the substrate where the topographical pattern provides a potential-energy landscape. 
To apply external driving forces to the colloidal clusters, the entire setup is tilted by an adjustable angle $\alpha$.
This results in a driving force per particle $F = mg \sin\alpha$, where $mg = 286$~fN is the buoyant weight of a colloid.
A detailed description of sample preparation and characterization is provided in Ref. \cite{cao2019np}.

\section{
Modeling, molecular dynamics, energy calculations and Fourier analysis.}
The particle-substrate potential energy of a particle at position \textbf{r} is a sum of infinite terms $V(\textbf{r}) = \sum_{n,m} V_\textrm{well}(|\textbf{r} - (n\textbf{b}_1 + m\textbf{b}_2)|)$, where $\textbf{b}_1$ and $\textbf{b}_2$ are the primitive vectors of the substrate, $n,m \in \mathbb{Z}$. $V_\textrm{well}(r)$ is a smooth approximation of the potential-energy profile for a colloid sphere located at a distance $r$ from the centre of a cylindrical well of radius $r_\textrm{M}$. We use $V_\textrm{well}(r)=-\epsilon$ for $r<r_\textrm{m}$; $V_\textrm{well}(r)=(-\epsilon/2)[\tanh( (w_\textrm{d}-\rho)/(\rho(1-\rho)) )+1]$ for $r_\textrm{m}<r<r_\textrm{M}$; $V_\textrm{well}(r)=0$ for $r>r_\textrm{M}$. 
Here, $\rho = (r-r_\textrm{m})/(r_\textrm{M}-r_\textrm{m})$. The parameters $w_\textrm{d} = 0.29$, $r_\textrm{m} = 0.6~\mu\rm{m}$ and $r_\textrm{M} = 2.0~\mu\rm{m}$ have been fitted to best replicate the experimental profile experienced by the $a = 4.45~\mu\rm{m}$ spheres. As in Ref. \cite{cao2019np}, we adopt an energy corrugation depth $\epsilon=105~\textrm{zJ} = 25.8 k_\textrm{B}T$  to best replicate experimental results. To reproduce the per-particle potential-energy landscape as in Fig.~1(h) we consider a cluster of particles fixed at positions $\textbf{r}_i = \textbf{r}_\textrm{c} + j_i \textbf{a}_1 + k_i \textbf{a}_2$, where $\textbf{r}_\textrm{c} = \sum_i \textbf{r}_i / N$ is the cluster's center-of-mass position, $\textbf{a}_1$ and $\textbf{a}_2$ are the primitive vectors of the colloidal lattice rotated at an angle $\theta_\textrm{o}$, and the set \{$j_i$, $k_i$\} for $i$ = 1, 2, 3,..., $N$ defines the cluster's size and shape. To simplify the formulation, here we have assumed $\sum_i j_i \textbf{a}_1 + k_i \textbf{a}_2 = 0$, i.e. the shape of the cluster is such that its center of mass is on a lattice point, initially located at the origin. The per-particle energy is then calculated as $U(\textbf{r}_\textrm{c}, \theta_\textrm{o}) = (1/N) \sum_i V(\textbf{r}_i)$.

To calculate the Fourier transform of $V(\textbf{r})$, we rewrite $V(\textbf{r})$ as :
\begin{equation}
\begin{split}
	    V(r) &= \sum_{n,m} V_\textrm{well}(|\textbf{r} - (n\textbf{b}_1 + m\textbf{b}_2)|) \\
             &= \iint_{\textbf{R}^2} \Sha(\textbf{r}')V_\textrm{well}(\textbf{r}-\textbf{r}') d\textbf{r}'.
\end{split}
\end{equation}
Here $\Sha(\textbf{r}') = \sum_{n,m} \delta_\textrm{Dirac}(n\textbf{b}_1 + m\textbf{b}_2 - \textbf{r}')$ is the 2D Dirac comb and $\delta_\textrm{Dirac}(\textbf{r})$ is the 2D Dirac delta function. We see that $V(\textbf{r})$ is the convolution of $\Sha(\textbf{r})$ and $V_\textrm{well}(\textbf{r})$. Therefore, according to convolution theory, the Fourier transform of $V(\textbf{r})$ is 
\begin{equation}
\begin{split}
    \tilde{V}(q)&\stackrel{\text{def}}{=}\mathcal{F}[V(\textbf{r})] \\ 
        &= \mathcal{F}[\Sha(\textbf{r})]\mathcal{F}[V_\textrm{well}(\textbf{r})] \\
        &= \sum_{n,m} \delta_\textrm{Dirac}[\textbf{q} - (n\bm{\beta}_1 + m\bm{\beta}_2)]\mathcal{F}[V_\textrm{well}(\textbf{r})] \\
        &= \Sha_{\textrm{q}}(\textbf{q}) \tilde{V}_\textrm{well}(\textbf{q}).
\end{split}           
\end{equation}			
Here $\Sha_{\textrm{q}}(\textbf{q})$ is the Dirac comb in reciprocal space, $\bm{\beta}_1$ and $\bm{\beta}_2$ are the reciprocal vectors of the lattice that satisfy $\bm{\beta}_i\cdot \textbf{b}_j=2\pi\delta_{ij}$ for $i,j=1,2$, $\tilde{V}_\textrm{well}(\textbf{q}) = \mathcal{F}[V_\textrm{well}(\textbf{r})]$ is the Fourier transform of $V_\textrm{well}(\textbf{r})$. Since $V_\textrm{well}(\textbf{r})$ is localized near \textbf{r} = 0 (i.e. the diameter of the cylindric well $|\Delta \textbf{r}| = 2r_\textrm{M} = 4~\mu\rm{m}$), $\tilde{V}_\textrm{well}(\textbf{q})$ will also be localized near \textbf{q} = 0, with a size $|\Delta\textbf{q}| \simeq 2\pi/|\Delta\textbf{r}|$ according to the uncertainty principle.
See also the numerically calculated $\tilde{V}_\textrm{well}(\textbf{q})$ in Fig.~S1.
Because of the $\tilde{V}_\textrm{well}(\textbf{q})$ factor, the Fourier component of $V(\textbf{r})$, namely $\delta_\textrm{Dirac}[\textbf{q} - (n\bm{\beta}_1 + m\bm{\beta}_2)]\tilde{V}_\textrm{well}(\textbf{q})$, will decay at large $|\textbf{q}|$.

\begin{figure}
  \centering
  \includegraphics[width=1.0\columnwidth]{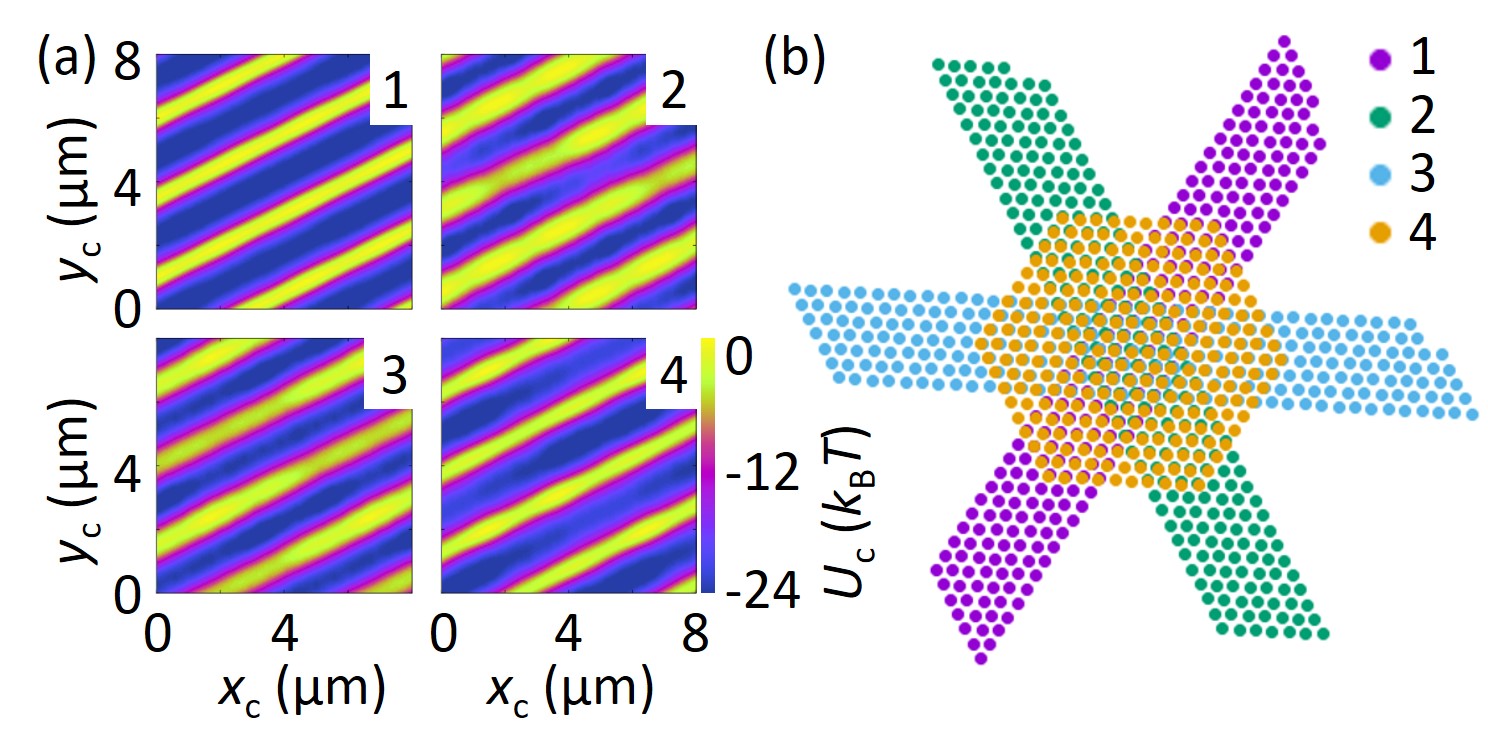}
\caption{\textbf{Effect of cluster shape on energy landscape.} (a) The calculated mean particle-substrate potential energy as a function of the center-of-mass position of each one of the four clusters in (b) on the $b = 5.0 ~\mu\textrm{m}$ square surface. The color scale and dimension is the same for the four panels. The energy landscape look quite similar, despite the quite different cluster shape and elongation. (b) Illustration of the four clusters of different shape/elongation. All clusters have the same spacing $a = 4.45 ~\mu\textrm{m}$, number of colloidal particles $N$ = 271, and fixed crystalline orientation $\theta_\textrm{o} = \textrm{-}3.43^{\circ}$.}
\label{fig2}
\end{figure}

\section{Results}
\subsection{Directional locking on square lattices}

Figure~1(b) and Supplementary Movie 1 \cite{supplementary} show an example of a measured center-of-mass trajectory (green line) of a triangularly packed colloidal cluster composed of $N=40$ particles driven across a square surface under an applied force \textbf{F} in the $x$ direction.
The substrate square pattern has a lattice spacing $b$ = 5.0 $\mu\rm{m}$ and is aligned with the $x-y$ coordinate system.
The angles $\varphi_\textrm{F}$, $\theta_\textrm{o}$ and $\theta_\textrm{d}$ relevant to the cluster motion are defined relative to the $x$-direction as shown in Fig.~1(b).
Here $\varphi_\textrm{F}$ is the angle of the driving force ($\varphi_\textrm{F}=0^{\circ}$ for \textbf{F} in the $x$ direction), $\theta_\textrm{o}$ and $\theta_\textrm{d}$ the cluster's orientation and moving direction respectively. The trajectory shows that $\theta_\textrm{d}$ is initially locked to $-26.6^{\circ}$ and then suddenly switches to 26.6$^{\circ}$. In between, the cluster's motion follows approximately the driving force (i.e. no directional locking).
Note that $\theta_\textrm{d}$ = $\pm 26.6^{\circ}$ is neither the direction of the driving force nor any of the high symmetry directions of the underlying substrate square lattice.
Fig.~1(c) shows a strong  correlation between $\theta_\textrm{d}$ and $\theta_\textrm{o}$.
When $\theta_\textrm{o}$ is locked to 3.4$^{\circ}$ ($-3.4^{\circ}$),  $\theta_\textrm{d}$ is locked to $-26.6^{\circ}$ (26.6$^{\circ}$). When $\theta_\textrm{o}$ deviates from 3.4$^{\circ}$ ($-3.4^{\circ}$), the cluster's trajectory is not directionally locked and follows roughly the driving direction. This correlation is also evident from the cluster motion in Movie 1. In general, the robustness of the directional and orientational locking strongly depends on the amplitude $F$ and the direction $\varphi_\textrm{F}$ of the driving force. This is shown in Fig.~1(d) and (e).
Fig.~1(d) shows the direction $\theta_\textrm{d}$ and  the corresponding orientation $\theta_\textrm{o}$ of sliding clusters as a function of the force direction $\varphi_\textrm{F}$. The force amplitude $F$ during the experiments is in the range 54--72~fN (to avoid that clusters come to rest, $F$ needs to be slightly increased when $\varphi_\textrm{F}$ deviates by more than $\sim 30^\circ$ from $\theta_\textrm{d}$).
Between $-10^{\circ} < \varphi_\textrm{F} < 60^{\circ}$, we identify a broad plateau where $\theta_\textrm{d}$ and $\theta_\textrm{o}$ remain robustly locked to 26.6$^{\circ}$ and $-3.43^{\circ}$ respectively. While at $\varphi_\textrm{F}$ $<$ $-10^{\circ}$ or at $\varphi_\textrm{F}$ $>$ 60$^{\circ}$, the clusters' moving direction tends to follow the driving  force as indicated by the dashed line ($\theta_\textrm{d}$ = $\varphi_\textrm{F}$), and the clusters' orientation either fluctuates irregularly around $-3.43^{\circ}$ or becomes locked to other orientations. Figure~1(e) shows the experimentally measured $\theta_\textrm{d}$ and $\theta_\textrm{o}$ as a function of the force amplitude $F$, keeping $\varphi_\textrm{F}$ = 0$^{\circ}$ fixed. We find a critical driving force (here, $F_\textrm{c} \approx$ 93~fN), above which the driven cluster dynamics follows the force direction, with small perturbations due to the underlying corrugation.

\begin{figure*}
  \centering
  \includegraphics[width=2.0\columnwidth]{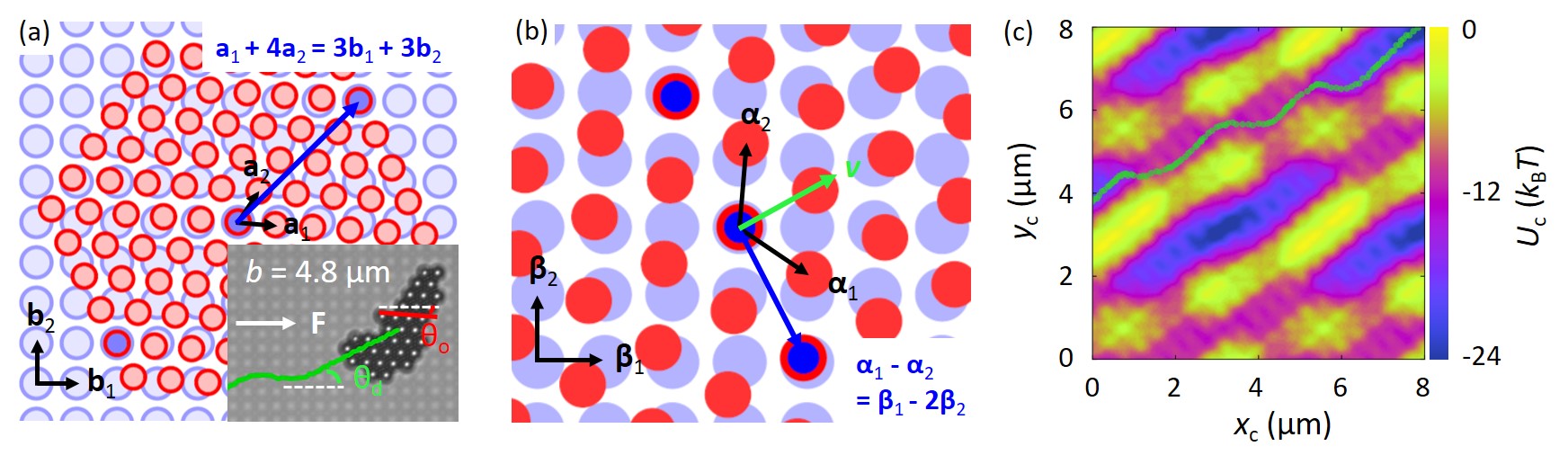}
\caption{\textbf{The directional locking behavior on a square surface lattice with smaller spacing $b = 4.8 ~\mu\rm{m}$.} (a) Illustration of the moir\'e pattern when a cluster at $\theta_\textrm{o}$ = $-4.1^{\circ}$ is in contact with the substrate. The blue arrow indicates the CLV for $(m_1, m_2, n_1, n_2)$ = (1, 4, 3, 3). (Inset) Center-of-mass trajectory (green) of an experimental cluster on the substrate, with $F$ = 89 fN and $\varphi_\textrm{F} = 0^{\circ}$. For most of the time in the trajectory, the cluster's direction is locked to $\theta_\textrm{d}$ = 26.6$^{\circ}$ and orientation $\theta_\textrm{o}$ = $-4.1^{\circ}$. (b) Moir\'e pattern of the reciprocal lattices for those in (a). The blue arrow indicates the reciprocal CLV for $(M_1, M_2, N_1, N_2) = (1, -1, 1, -2)$. (c) Calculated potential energy landscape for the cluster in (a). The green dotted line is the center-of-mass trajectory of the cluster in (a) inset for a periodic of 36 s, data points are taken at 0.33 s interval.}
\label{fig3}
\end{figure*}

For a triangular cluster driven across a triangular substrate \cite{cao2019np}, orientational locking originates from a (nearly) commensurate matching of the moir\'e pattern of the contacting colloidal and substrate lattices, i.e.  
\begin{equation} \label{eq1}
\textbf{R}_\textrm{M} \stackrel{\text{def}}{=} m_1\textbf{a}_1 + m_2\textbf{a}_2 = n_1\textbf{b}_1 + n_2\textbf{b}_2.
\end{equation} 
Here ($\textbf{a}_1$, $\textbf{a}_2$) are the primitive vectors of the colloidal cluster, ($\textbf{b}_1$, $\textbf{b}_2$) the primitive vectors of the periodic surface, and $m_1, m_2, n_1, n_2$ are suitable integers such that $\textbf{R}_\textrm{M}$ is the smallest coincidence lattice vector (CLV) common to both lattices. Being immediately obvious for same-symmetry lattices in contact, in the following we show that this expression still applies to the current situation of a triangular cluster on a square substrate lattice. Figure~1(f) schematically shows a hexagon-shaped colloidal cluster (red particles) oriented at $\theta_\textrm{o}$ = $-3.4^{\circ}$ on top of a square-lattice pattern (light blue dots representing wells). As highlighted by the blue particles, there is nearly commensurate matching of the two lattices in our  experiments. As indicated by the blue arrow, this moir\'e pattern is characterized by the CLV $5\textbf{a}_1 - 5\textbf{a}_2 \approx 2\textbf{b}_1 - 4\textbf{b}_2$, i.e. $(m_1, m_2, n_1, n_2) = (5, -5, 2, -4)$. For a triangle-square contact, equation~(\ref{eq1}) requires that the lattice-spacing ratio satisfies $a/b = \lambda_\textrm{R}$, where $\lambda_\textrm{R} = \sqrt{(n_1^2 + n_2^2)/(m_1^2 + m_2^2 + m_1m_2 )}$. This is not strictly satisfied for our experiments with $a/b$ = 4.45/5.0 and $(m_1, m_2, n_1, n_2) = (5, -5, 2, -4)$. However, the deviation $\delta_\textrm{R} = |\delta\textbf{R}_\textrm{M}| / |\textbf{R}_\textrm{M}| = |5\textbf{a}_1 - 5\textbf{a}_2 - (2\textbf{b}_1 - 4\textbf{b}_2)| / |2\textbf{b}_1 - 4\textbf{b}_2| = |(1/\lambda_\textrm{R}) a/b - 1|$ = 0.5\% is small enough that orientational locking can be observed in our experiments  for clusters up to $N \sim 200$ with various shapes (see Supplementary Movie 2). The orientation of the cluster as derived from equation~(\ref{eq1}) is $\theta_\textrm{o} = \arctan(n_2/n_1) -  \arctan[(\sqrt{3}/2) m_2 / (m_1+m_2/2)]$. With the integers $(m_1, m_2, n_1, n_2) = (5, -5, 2, -4)$, the calculated $\theta_\textrm{o} = -3.43^{\circ}$ agrees well with experimental observations. Note that, due to the special choice of $a/b$ = 4.45/5.0, in Fig.~1(f) there are two additional well-matched lattice vectors $5\textbf{a}_1 - \textbf{a}_2 \approx 4\textbf{b}_1 - \textbf{b}_2$ and $4\textbf{a}_2 \approx 2\textbf{b}_1 + 3\textbf{b}_2$. These two extra well-matched lattice vectors suggest the additional stability of the locking orientation at $\theta_\textrm{o} = -3.43^{\circ}$.

Given a fixed cluster orientation $\theta_\textrm{o}$, its direction of motion depends on the driving force \textbf{F} as well as the interaction energy per particle $U(\textbf{r}_\textrm{c}, \theta_\textrm{o}) = (1/N) \sum_j V(\textbf{r}_j)$ between the cluster and the substrate. Here $\textbf{r}_\textrm{c} = (x_\textrm{c}, y_\textrm{c}) = (1/N) \sum_j \textbf{r}_j$ is the cluster's center of mass position, $V(\textbf{r})$ is the particle-substrate interaction energy and $N$ is the number of particles in the cluster. As derived in ref.~\cite{cao2019np}, 
\begin{equation} \label{eq2}
U(\textbf{r}_\textrm{c}, \theta_\textrm{o}) \cong \sum_\textbf{Q} \tilde{V}(\textbf{Q}) \exp(i\textbf{Q}\cdot\textbf{r}_\textrm{c}).		       	
\end{equation} 
Here $i$ is the imaginary unit, $\tilde{V}(\textbf{q})$ is the Fourier transform of $V(\textbf{r})$, and \textbf{Q} is a reciprocal lattice vector of the cluster. Since $V(\textbf{r})$ is a periodic potential, $\tilde{V}(\textbf{Q})$ is nonzero only when \textbf{Q} is also a reciprocal lattice vector of the substrate, i.e.\ \textbf{Q} is a CLV in the reciprocal space of both contacting lattices, which satisfies:
\begin{equation} \label{eq3}
\textbf{Q} \stackrel{\text{def}}{=} M_1 \bm{\alpha}_1 + M_2 \bm{\alpha}_2 = N_1 \bm{\beta}_1 + N_2 \bm{\beta}_2.
\end{equation} 
Here $M_1, M_2, N_1, N_2$ are integers, $\bm{\alpha}_1, \bm{\alpha}_2, \bm{\beta}_1, \bm{\beta}_2$ are the reciprocal primitive vectors that satisfy $\bm{\alpha}_i\cdot\textbf{a}_j=2\pi\delta_{ij}$ and $\bm{\beta}_i\cdot\textbf{b}_j=2\pi\delta_{ij}$ for $i,j=1,2$.
Since $\tilde{V}(\textbf{q})$ decays rapidly as $|\textbf{q}|$ (and therefore the order of the Fourier component) increases, for simplicity we consider only the dominating lowest-order Fourier component determined by the shortest reciprocal CLV, in this case $\pm\textbf{Q}_\textrm{M}$.
Then equation~(\ref{eq2}) approximately simplifies to $U(\textbf{r}_\textrm{c}, \theta_\textrm{o}) \approx 2\tilde{V}(\textbf{Q}_\textrm{M}) \cos(\textbf{Q}_\textrm{M}\cdot\textbf{r}_\textrm{c})$.
Note that $\tilde{V}(\textbf{Q}_\textrm{M})$ is a real quantity since $V(\textbf{r})$ has inversion symmetry, $U(\textbf{r}_\textrm{c}, \theta_\textrm{o})$ is a two-dimensional sinusoidal wave with wave vector $\textbf{Q}_\textrm{M}$.
The cluster's direction of motion then follows a trough of this sinusoidal wave, which is a straight line perpendicular to $\textbf{Q}_\textrm{M}$.
Fig.~1(g) depicts the reciprocal lattices for the real-space lattices of Fig.~1(f). Here the shortest CLV $\textbf{Q}_\textrm{M} = \bm{\alpha}_1 - \bm{\alpha}_2 \approx \bm{\beta}_1 - 2\bm{\beta}_2$ (blue line) is clearly seen. The direction perpendicular to $\textbf{Q}_\textrm{M}$ gives $\theta_\textrm{d} = \arctan(-N_1/N_2) = 26.6^{\circ}$ which agrees perfectly with the observed locked sliding directions. Indeed, as illustrated in Fig.~1(h), the numerically calculated $U(\textbf{r}_\textrm{c}, \theta_\textrm{o}=-3.43^{\circ})$ for the cluster of Fig.~1(f) reveals an approximately sinusoidal modulation with a sequence of low-energy troughs or corridors oriented along the 26.6$^{\circ}$ direction. In addition, the potential energy landscape agrees very well with the experimental cluster's trajectory (green dotted line), even though the experimental cluster does not have the same size and shape as that of Fig.~1(f).

Directional locking occurs as long as the perpendicular component of the driving force $\textbf{F}_{\perp}$ is insufficient to overcome the barrier between adjacent corridors, i.e. $F_{\perp} < F_{\textrm{c}\perp}$. Given $F_\textrm{c}\approx$ 93~fN at $\varphi_\textrm{F} = 0^{\circ}$ in Fig.~1(e), we obtain $F_{\textrm{c}\perp} = F_{\textrm{c}} \sin(\theta_\textrm{d} - \varphi_\textrm{F}) \approx$ 42~fN.
$F_{\textrm{c}\perp}$ can also be evaluated from Fig.~1(d) where directional locking disappears beyond a critical angle $\varphi_\textrm{F,c}\approx -10^{\circ}$ at F = 72~fN, i.e. $F_{\textrm{c}\perp} = F \sin(\theta_\textrm{d} - \varphi_\textrm{F,c}) \approx$ 43~fN. 
Note that the cluster size and shape in experiments differs from one cluster to another and is not necessarily the same as that in Fig.~1(f) which is used for calculating the $U(\textbf{r}_\textrm{c}, \theta_\textrm{o})$.
This does not affect our results here since the cluster's shape has only little influence on the calculated $U(\textbf{r}_\textrm{c}, \theta_\textrm{o})$ as long as the cluster's orientation remains the same and the cluster size is similar. This is clearly shown in Fig.~2 where we plot the $U(\textbf{r}_\textrm{c}, \theta_\textrm{o}=3.43^{\circ})$ of clusters with different shapes.

Moir\'e patterns in reciprocal and real space, as characterized by the CLVs $\textbf{Q}_\textrm{M}$ and $\textbf{R}_\textrm{M}$ respectively, are strictly related to each other.
Since directional locking is determined by the moir\'e pattern in reciprocal space, the knowledge of how to obtain $\textbf{Q}_\textrm{M}$ from $\textbf{R}_\textrm{M}$ allows us to calculate $\theta_\textrm{d}$.
For a triangular cluster on a square lattice, the relation is given by
\begin{equation} \label{eq4}
\begin{split}
      M_1 &= \frac{(2m_1+m_2)(n_1^2+n_2^2)}{K}, \\
      M_2 &= \frac{(m_1+2m_2)(n_1^2+n_2^2)}{K},	\\
      N_1 &= \frac{2n_1(m_1^2+m_2^2+m_1m_2)}{K},	\\
      N_2 &= \frac{2n_2(m_1^2+m_2^2+m_1m_2)}{K}.
\end{split}
\end{equation} 
See Appendix A for the derivation of equations~(\ref{eq4}).
Here $K$ is an integer, selected in order to make $M_1, M_2, N_1, N_2$ coprime integers.
This expression is more involved compared with the simple relation $(M_1, M_2, N_1, N_2) = (n_1, -n_2, m_1, -m_2)$ holding for two triangular lattices in contact \cite{cao2019np}.
For the case $(m_1, m_2, n_1, n_2) = (5, -5, 2, -4)$ of our experiments in Fig.~1(f), the value of $K$ is 100.
This gives $(M_1, M_2, N_1, N_2) = (1, -1, 1, -2)$, which is exactly the $\textbf{Q}_\textrm{M} = \bm{\alpha}_1 - \bm{\alpha}_2 \approx \bm{\beta}_1 - 2\bm{\beta}_2$ shown in Fig.~1(g). Similar to equation~(\ref{eq1}), equation~(\ref{eq3}) requires that the lattice-spacing ratio in reciprocal space satisfies $|\bm{\beta}_1|/|\bm{\alpha}_1| = \lambda_\textrm{Q}$, where $\lambda_\textrm{Q} = \sqrt{(M_1^2 + M_2^2 - M_1M_2 ) / (N_1^2 + N_2^2)} $.
This is not strictly satisfied for our experiments with $|\bm{\beta}_1|/|\bm{\alpha}_1| = (\sqrt{3}/2)a/b = 0.771$ and $(M_1, M_2, N_1, N_2) = (1, -1, 1, -2)$. The deviation $\delta_\textrm{Q} = |(1/\lambda_\textrm{Q}) |\bm{\beta}_1|/|\bm{\alpha}_1| - 1| = \delta_\textrm{R}$ as inherited from equation~(\ref{eq4}).

\begin{figure*}
  \centering
  \includegraphics[width=2.0\columnwidth]{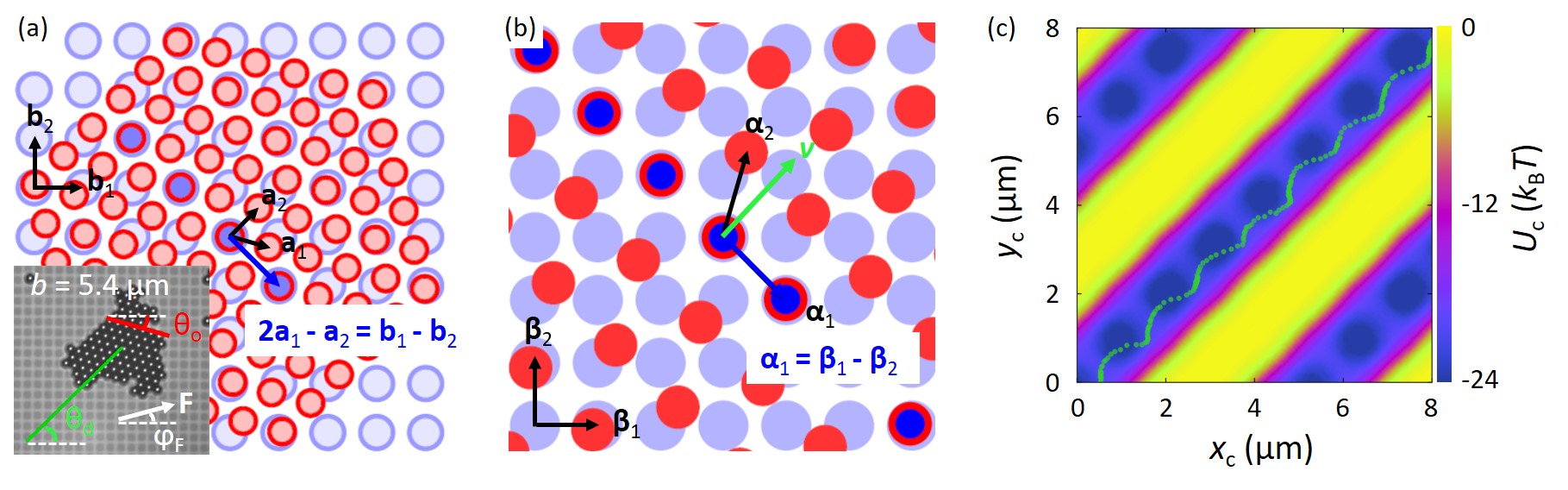}
\caption{\textbf{Orientational and directional locking of colloidal clusters on $b = 5.4~\mu\textrm{m}$ square surfaces.} (a) The moir\'e pattern of an $a = 4.45 ~\mu\textrm{m}$ cluster with $\theta_\textrm{o} = \textrm{-}15.0^{\circ}$ on top of a $b = 5.4~\mu\textrm{m}$ square lattice. The blue arrow indicates $\textbf{R}_\textrm{m} = 2\textbf{a}_1 - \textbf{a}_2 \approx \textbf{b}_1 - \textbf{b}_2$, i.e. $(m_1, m_2, n_1, n_2)$ = (2, -1, 1, -1). (inset) Trajectory of an experimental cluster on $b = 5.4~\mu\textrm{m}$ square lattice, with $F$ = 89 fN and $\varphi_\textrm{F} = 15.0^{\circ}$. The cluster's direction of motion is locked at $\theta_\textrm{d} = 45.0^{\circ}$ and the orientation at $\theta_\textrm{o} = -15.0^{\circ}$. (b) The moir\'e pattern of the corresponding reciprocal lattices. The blue arrow indicates $\textbf{Q}_\textrm{m} = \bm{\alpha}_1 \approx \bm{\beta}_1 - \bm{\beta}_2$, i.e. $(M_1, M_2, N_1, N_2)$ = (1, 0, 1, -1). $(M_1, M_2, N_1, N_2)$ and $(m_1, m_2, n_1, n_2)$ fulfill the relation in equation~(\ref{eq4}) with $K = 6$. (c) The calculated potential energy landscape for the cluster in (a) on the $b = 5.4~\mu\textrm{m}$ square lattice. The low-energy corridor along the 45.0$^{\circ}$ direction is very robust. The green dotted line is the center-of-mass trajectory of the cluster in (a) inset for a period of 60 s, data points are taken at 0.33 s interval.
}
\label{fig4}
\end{figure*}

\begin{figure*}
  \centering
  \includegraphics[width=2.0\columnwidth]{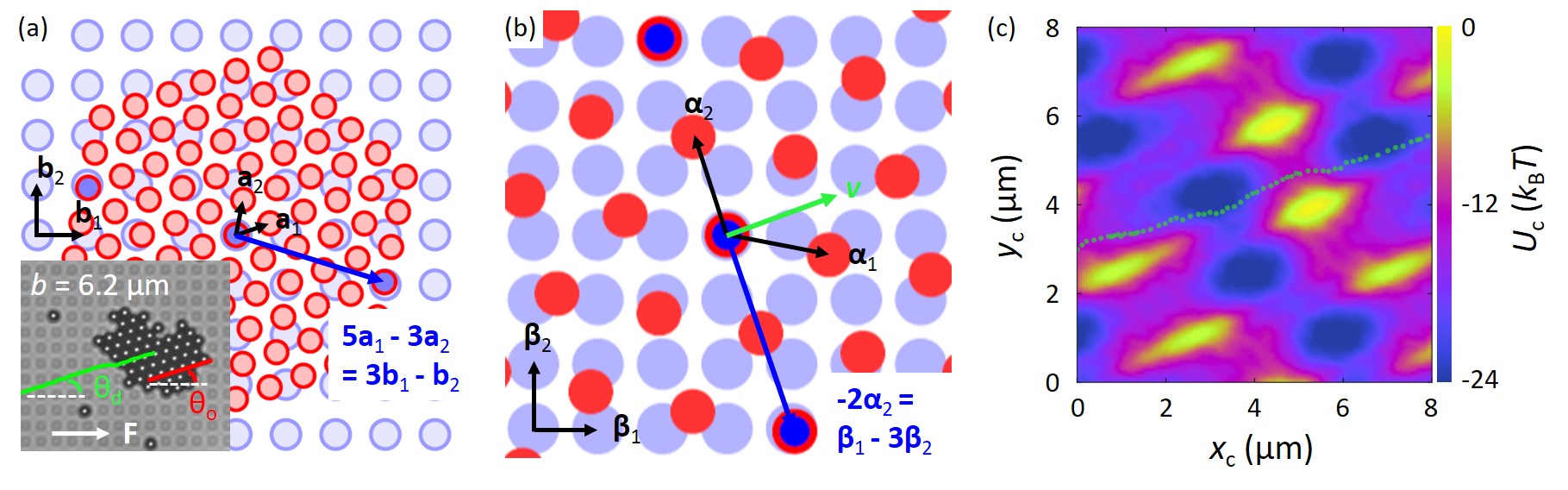}
\caption{\textbf{The directional-locking behavior on a square surface lattice with spacing $b = 6.2~\mu\textrm{m}$.} (a) The moir\'e pattern obtained when a cluster at  $\theta_\textrm{o} = 18.2^{\circ}$ is in contact with the substrate. The blue arrow indicates the CLV for $(m_1, m_2, n_1, n_2)$ = (5, -3, 3, -1). (Inset) Center-of-mass trajectory (green) of an experimental cluster on the substrate, with $F$ = 89 fN and $\varphi_\textrm{F} = 0^{\circ}$. For most of the time in the trajectory, the cluster's direction is locked to $\theta_\textrm{d} = 18.4^{\circ}$ and orientation $\theta_\textrm{o} = 18.2^{\circ}$. (b) Moir\'e pattern of the reciprocal lattices relative to those in (a). The blue arrow indicates the approximate reciprocal CLV obtained for $(M_1, M_2, N_1, N_2)$ = (0, -2, 1, -3). $(M_1, M_2, N_1, N_2)$ and $(m_1, m_2, n_1, n_2)$ do not fulfill the relation in equation~(\ref{eq4}). (c) Calculated potential energy landscape for the cluster in (a). Part of the "barrier walls" between adjacent low-energy corridors are "etched away". The green dotted line is the center-of-mass trajectory of the cluster in (a) inset for a period of 20 s, data points are taken at 0.33 s interval.
}
\label{fig5}
\end{figure*}

The connection between the smallest CLV in real space and in reciprocal space, as characterized by equation~(\ref{eq4}), could become lost for the triangle-square contact when the deviation $\delta_\textrm{R}$ or $\delta_\textrm{Q}$ from a perfect matching becomes large.
This is shown in Fig.~3, where we slightly shrink the substrate lattice spacing from $b = 5.0~\mu\rm{m}$ to $b = 4.8~\mu\rm{m}$ in experiments (the colloidal cluster's spacing remains $a = 4.45~\mu\rm{m}$).
As shown in Fig.~3(a) (inset) and Supplementary Movie 3, directional/orientational locking is still observed after we have slightly shrunk the substrate lattice.
However, the orientation of the cluster changes slightly to $\theta_\textrm{o} = -4.1^{\circ}$, while the direction of motion of the colloidal cluster remains at $\theta_\textrm{d}$ = 26.6$^{\circ}$. Correspondingly, the smallest CLV in real space is changed to $\textbf{R}_\textrm{M}^* = \textbf{a}_1 + 4\textbf{a}_2 \approx 3\textbf{b}_1 + 3\textbf{b}_2$ as shown in Fig.~3(a), while the CLV in reciprocal space remains to be $\textbf{Q}_\textrm{M} = \bm{\alpha}_1 - \bm{\alpha}_2 \approx \bm{\beta}_1 - 2\bm{\beta}_2$ as shown in Fig.~3(b).
The reason for the change of CLV in real space is, when we change the substrate lattice spacing from $b$ = 5.0 $\mu\rm{m}$ to $b$ = 4.8 $\mu\rm{m}$, the deviation $\delta_\textrm{R}$ changes from 0.5\% to 3.7\% for the CLV $\textbf{R}_\textrm{M} = 5\textbf{a}_1 - 5\textbf{a}_2 \approx 2\textbf{b}_1 - 4\textbf{b}_2$, which is very large. Therefore, the cluster tends to slightly rotate toward the much-better-matched moir\'e pattern characterized by the CLV $\textbf{R}_\textrm{M}^* = \textbf{a}_1 + 4\textbf{a}_2 \approx 3\textbf{b}_1 + 3\textbf{b}_2$ whose $\delta_\textrm{R}$ is only 0.1\% for the $b = 4.8 ~\mu\rm{m}$ substrate.
After the slight rotation, the best matched and relatively small-sized CLV in reciprocal space remains $\textbf{Q}_\textrm{M} = \bm{\alpha}_1 - \bm{\alpha}_2 \approx \bm{\beta}_1 - 2\bm{\beta}_2$ with $\delta_\textrm{Q} = |\bm{\beta}_1 - 2\bm{\beta}_2 - (\bm{\alpha}_1 - \bm{\alpha}_2)| / |\bm{\alpha}_1 - \bm{\alpha}_2| = 3.8\%$.
Clearly, $\textbf{R}_\textrm{M}^*$ and $\textbf{Q}_\textrm{M}$ can no longer be connected with equation~(\ref{eq4}). One consequence of this poorer matching at $b = 4.8~\mu\rm{m}$ is that the orientational and directional locking become much less robust compared to those of Fig.~1, where the connection between the smallest CLVs in real space and in reciprocal space is exactly governed by equation~(\ref{eq4}). This is confirmed by the calculated potential-energy landscape shown in Fig.~3(c), where the barriers between the low-energy troughs are substantially fragmented and distorted compared with Fig.~1(h). Accordingly, the cluster's trajectory shows a strong zigzag behavior.
Besides the $b = 5.0~\mu\rm{m}$ and $b = 4.8~\mu\rm{m}$ square lattice, we have also found directional locking on $b = 5.4~\mu\rm{m}$ and $b = 6.2~\mu\rm{m}$ square lattice, which is shown in Fig.~4/Movie 4 and Fig.~5/Movie 5 respectively and can be explained with similar geometrical argument with that of the $b = 5.0~\mu\rm{m}$ and $b = 4.8~\mu\rm{m}$ cases.  

\begin{figure*}
  \centering
  \includegraphics[width=2.0\columnwidth]{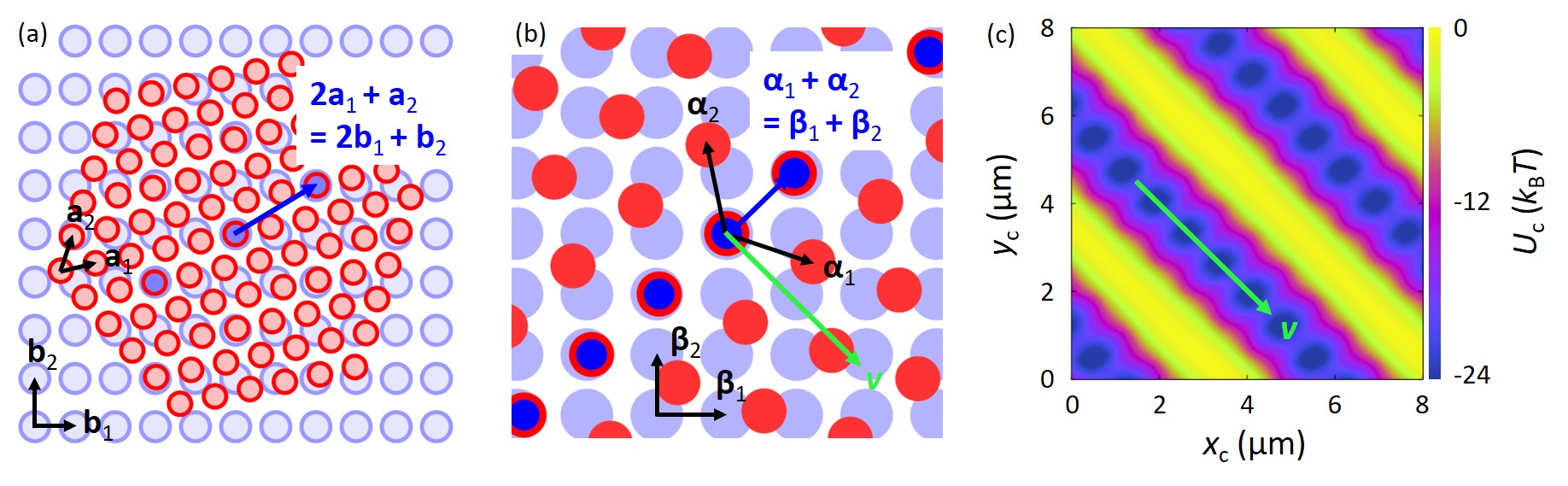}
\caption{\textbf{Orientational and directional locking of a simulated cluster on a rectangular lattice with $|\textbf{b}_1| = 5.0 ~\mu\rm{m}$ and $|\textbf{b}_2| = 6.0 ~\mu\rm{m}$.} (a) Illustration of the moir\'e pattern of a cluster on top of the rectangular lattice. The angle between $\textbf{a}_1$ and $\textbf{b}_1$ (i.e. $\theta_\textrm{o}$) is 11.9$^{\circ}$. $a = |\textbf{a}_1| = |\textbf{a}_2| = 4.45 \mu\rm{m}$. A CLV $\textbf{R}_\textrm{M} = 2\textbf{a}_1 + \textbf{a}_2 \approx 2\textbf{b}_1 + \textbf{b}_2$ is clearly seen. (b) The moir\'e pattern in reciprocal space for the two lattices in (a).  A CLV  $\textbf{Q}_\textrm{M} = \bm{\alpha}_1 + \bm{\alpha}_2 \approx \bm{\beta}_1 + \bm{\beta}_2$ is clearly seen. (c) The numerically calculated potential energy $U(\textbf{r}_\textrm{c}, \theta_\textrm{o})$ for the cluster in (a). }
\label{fig6}
\end{figure*}

\subsection{Directional locking on a rectangular lattice}
An important indication from equation~(\ref{eq2}) is that directional locking can in principle be observed for a cluster with arbitrary lattice symmetry sliding across another arbitrary lattice, as long as a CLV $\textbf{Q}$ exists in reciprocal space, and the relative Fourier component $\tilde{V}(\textbf{Q})$ of the particle-substrate potential energy is nonzero.
This is showcased by our numerically calculated $U(\textbf{r}_\textrm{c}, \theta_\textrm{o})$ in Fig.~6(c) for a triangular cluster sliding across a rectangular lattice.
The moir\'e pattern of the two contacting lattices in real space and in reciprocal space is shown in Fig.~6(a) and 6(b) respectively, with $\textbf{R}_\textrm{M} = 2\textbf{a}_1 + \textbf{a}_2 \approx 2\textbf{b}_1 + \textbf{b}_2$ and $\textbf{Q}_\textrm{M} = \bm{\alpha}_1 + \bm{\alpha}_2 \approx \bm{\beta}_1 + \bm{\beta}_2$.
The integer coefficients of equation~(\ref{eq3}) are related to those of equation~(\ref{eq1}) by
\begin{equation} \label{eq5}
\begin{split}
      M_1 &= \epsilon n_1 + \eta \lambda_\textrm{b} n_2, \\
      M_2 &= [\frac{\eta\sin\phi-\epsilon\sin(\phi-\psi)}{\sin\psi}n_1 \\
      	  &+ \frac{-\epsilon\sin\phi + \eta\sin(\phi+\psi)}{\sin\psi}\lambda_\textrm{b}n_2]\lambda_\textrm{a},	\\
      N_1 &= \epsilon m_1 + \frac{\eta\sin\phi-\epsilon\sin(\phi-\psi)}{\sin\psi}\lambda_\textrm{a} m_2,	\\
      N_2 &= [\eta m_1 + \frac{-\epsilon\sin\phi + \eta\sin(\phi+\psi)}{\sin\psi}\lambda_\textrm{a}m_2]\lambda_\textrm{b}.
\end{split}
\end{equation} 
Here $\lambda_\textrm{a} = |\textbf{a}_2|/|\textbf{a}_1| =1$, $\lambda_\textrm{b} = |\textbf{b}_2|/|\textbf{b}_1| =1.2$, $\phi$=60$^{\circ}$ and $\psi$=90$^{\circ}$ are angles between $\textbf{a}_1$ and $\textbf{a}_2$ and between $\textbf{b}_1$ and $\textbf{b}_2$ respectively, $\epsilon=0.4328$ and $\eta=0.2403$ are real numbers such that $M_1, M_2, N_1, N_2$ are coprime integers.
The derivation of equation~(\ref{eq5}) is detailed in Appendix B, where we provide an implicit relation between $\textbf{R}_\textrm{M}$ and $\textbf{Q}_\textrm{M}$ for two arbitrary 2D lattices in contact.
Note that the calculated $(M_1, M_2, N_1, N_2)$ from equation~(\ref{eq5}) is (1, 1, 0.98, 1.05), which is not exactly (1, 1, 1, 1).
Such deviations will generally exist when non-equilateral lattices (such as the rectangular lattices) are involved. However, directional locking can be observed as long as equation~(\ref{eq1}) and equation~(\ref{eq3}) are nearly satisfied.

\begin{figure*}
  \centering
  \includegraphics[width=2.0\columnwidth]{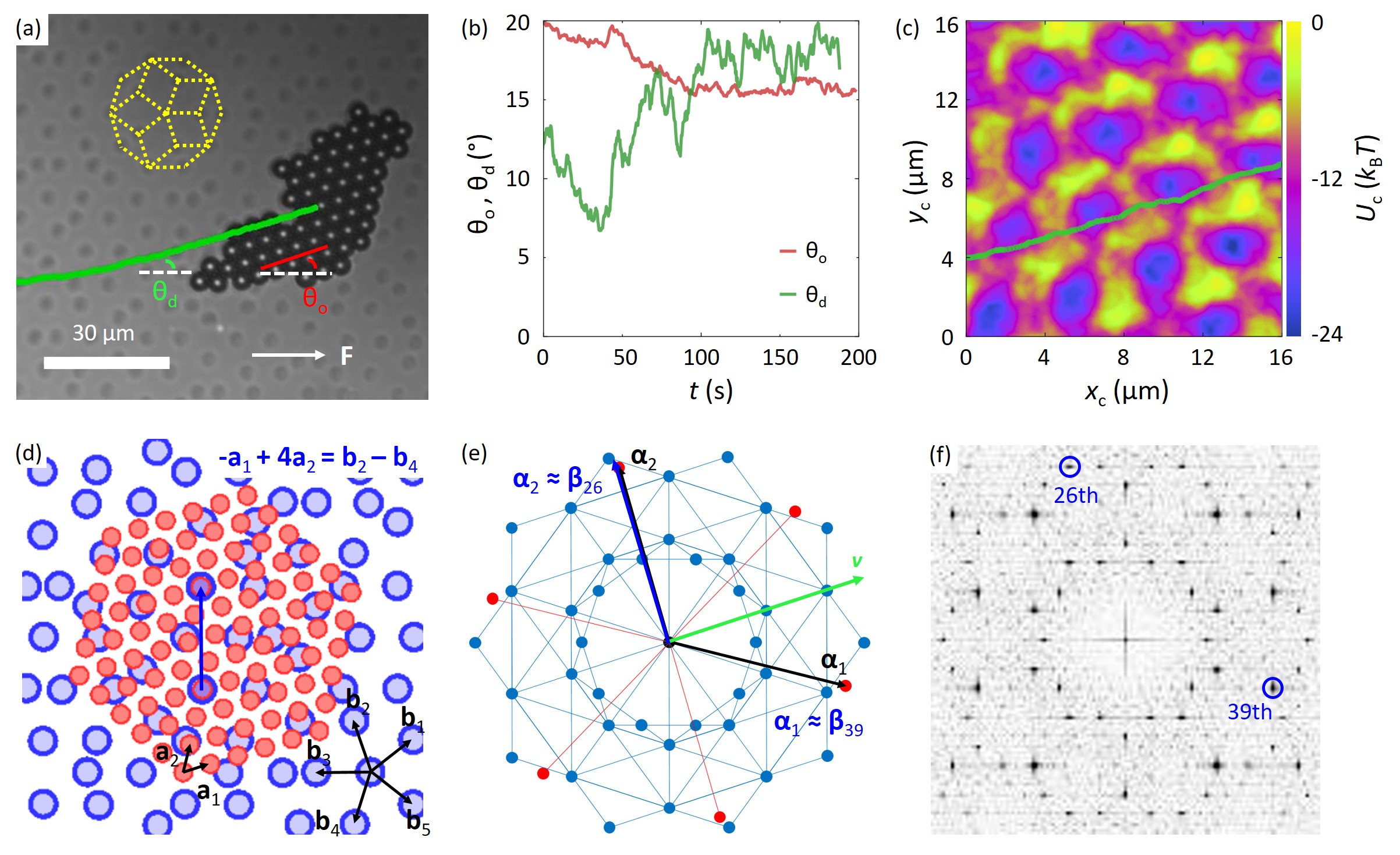}
\caption{\textbf{Orientational and directional locking on quasiperiodic surfaces.} (a) A colloidal cluster in experiments showing directional locking on a quasiperiodic surface.
Green line is the cluster's center of mass trajectory under $F$ = 89 fN and $\varphi_\textrm{F} = 0^{\circ}$.
Five fat rhombi with internal angles 72$^{\circ}$/108$^{\circ}$ and five slim rhombus with internal angles 36$^{\circ}$/144$^{\circ}$ are highlighted with dashed lines. Each line connects the centers of two substrate wells.
The rhombi's side length is $b$ = 8.51 $\mu\rm{m}$.
(b) The orientation of the cluster $\theta_\textrm{o}$ and its direction of motion $\theta_\textrm{d}$ as a function of time $t$. 
(c) The calculated mean particle-substrate potential energy of the cluster in (d) as a function of its center-of-mass position $x_\textrm{c}$ and $y_\textrm{c}$, for energetically optimal $\theta_\textrm{o}$ = 16.0$^{\circ}$. 
The potential-energy landscape exhibits a low-energy corridor along the $\sim 18^{\circ}$ direction. The color code is the same as that in Fig.~1(h). The green (densely dotted) line is the center-of-mass trajectory of the cluster in (a,b) for a period of time from $t=150$ s to $t=190$ s, data points are taken at 0.33 s interval.
(d) The moir\'e pattern of an $a$ = 4.45 $\mu\rm{m}$ cluster with $\theta_\textrm{o}$ = 16.0$^{\circ}$ on top of the quasiperiodic lattice. The blue arrow indicates $-\textbf{a}_1 + 4\textbf{a}_2 \approx \textbf{b}_2 - \textbf{b}_4$. Note that only four ``primitive" vectors $\textbf{b}_1, \textbf{b}_2, \textbf{b}_3, \textbf{b}_4, [\textbf{b}_5 = -(\textbf{b}_1 + \textbf{b}_2 + \textbf{b}_3 + \textbf{b}_4)]$ are needed to describe the quasiperiodic lattice. $|\textbf{b}_1| = |\textbf{b}_2| = |\textbf{b}_3| = |\textbf{b}_4| = |\textbf{b}_5| = b$.
(e) The moir\'e pattern of the corresponding reciprocal lattices in (d).
The blue arrow indicates $\bm{\alpha}_2 \approx \bm{\beta}_{26}$, i.e. $\bm{\alpha}_2$ overlaps with the $26^\textrm{th}$ peak in the scattering pattern of the quasiperiodic structure.
(f) The scattering pattern of the quasiperiodic surface in our experiments; the $26^\textrm{th}$ and $39^\textrm{th}$ peaks are highlighted.
The image represents the peak intensity in log scale.
}
\label{fig7}
\end{figure*}

\subsection{Directional locking on a quasiperiodic lattice}
Finally, besides the orientational and directional locking of colloidal clusters on periodic surfaces, we also find directional locking on a quasiperiodic substrate.
This is illustrated by the data shown in Fig.~7(a) and Supplementary Movie 6.
Here, the quasiperiodic substrate is fabricated by arranging the substrate wells at the vertices of fat rhombus and slim rhombus motifs alternating following the rules of a Penrose-like tiling, which is highlighted by the dashed lines in Fig.~7(a). See Fig.~S2 for more details of the quasiperiodic structure.
The orientation $\theta_\textrm{o}$ and direction of motion $\theta_\textrm{d}$ of the cluster as a function of time $t$ is reported in Fig.~7(b).
After a transient period, the cluster's orientation is locked to $\theta_\textrm{o} \approx 16^{\circ}$, while the direction of motion is locked to $\theta_\textrm{d} \approx 16^{\circ}$--$18^{\circ}$.
Figure~7(c) shows $U(\textbf{r}_\textrm{c}, \theta_\textrm{o})$ for a cluster on the quasiperiodic substrate, a sequence of clear but relatively weak low-energy troughs along the $16^{\circ}$--$18^{\circ}$ direction is revealed.
Note that, directional locking has also been observed when interacting colloidal particles are driving through quasiperiodic substrates, where the direction of colloidal particles are locked to the symmetry directions of the quasiperiodic structure \cite{reichhardt2011prl,bohlein2012prl}. This is different from the directional locking reported in this work, where the locking direction is determined by the commensurate matching of the two lattices in contact and is not necessarily the substrate symmetry direction. 

Similar to the periodic surface, we observe an approximate CLV $-\textbf{a}_1 + 4\textbf{a}_2 \approx \textbf{b}_2 - \textbf{b}_4$ in the superimposition of the colloidal cluster and the quasiperiodic surface as shown in Fig.~7(d).
This CLV determines the orientation of the colloidal cluster to be $\theta_\textrm{o}$ = 16.1$^{\circ}$, in agreement with our experimental observations.
Contrary to the case of a periodic substrate, the reciprocal space (i.e. the Fourier transform) of a quasiperiodic lattice is a discrete set of points which do not form a lattice.
We consider the brightest points near the origin which corresponds to low-order Fourier components.
These points form a decagonal structure.
The pattern shown in Fig.~7(e,f) is obtained by the Fourier transform of a 250 $\mu\rm{m}$ $\times$ 250 $\mu\rm{m}$ quasiperiodic region.
By comparing the position of these points with that of the reciprocal lattice points of the colloid cluster in Fig.~7(e), we observe a well-matched lattice vector $\bm{\alpha}_2 \approx \bm{\beta}_{26}$, where $\bm{\beta}_{26}$ is the $26^\textrm{th}$ peak in the scattering pattern of the quasiperiodic surface, as shown in Fig.~7(f).
In addition, the lattice vector $\bm{\alpha}_1$ also matches approximately the $39^\textrm{th}$ peak $\bm{\beta}_{39}$, leading to the hexagonal-like potential energy landscape in Fig.~7(c).
Considering that $|\bm{\alpha}_2| = 2\pi / (a\cos 30^{\circ}) = 1.63 \mu\rm{m}^{-1}$ pointing toward the $-74^{\circ}$ direction and $|\bm{\beta}_{26}| = (4\pi/b)\tau^3/(1+\tau^2)  = 1.73 \mu\rm{m}^{-1}$ ($\tau=1.618\dots$ is the golden ratio) pointing toward the $-72^{\circ}$ direction, the matching $\bm{\alpha}_2 \approx \bm{\beta}_{26}$ is significantly imperfect.
This makes the directional locking behavior rather weak. This explains why $\theta_\textrm{d}$ fluctuates between $90^{\circ}-74^{\circ}=16^{\circ}$ and $90^{\circ}-72^{\circ}=18^{\circ}$, as observed in experiments.

From the above matching conditions, we see that equation~(\ref{eq2}) is also relevant for quasiperiodic structures even though it is originally derived for periodic ones.
In fact, equation~(\ref{eq2}) is valid whenever the particle-substrate potential $V(\textbf{r})$ is periodic or quasiperiodic, i.e. when it has a well defined Fourier transform of the form $\tilde{V}(\textbf{q}) = \mathcal{F}[V(\textbf{r})] = \sum_\textbf{G} \tilde{V}(\textbf{G}) \delta_{\textbf{q}, \textbf{G}}$, with $\textbf{G} \in \bm{\Sigma}$. If $V(\textbf{r})$ is periodic, $\bm{\Sigma}$ is a lattice; if $V(\textbf{r})$ is quasiperiodic then $\bm{\Sigma}$, while not a lattice, is still a discrete set.
To unify notation for periodic and quasiperiodic substrate, we use the definition $\mathcal{F}[V(\textbf{r})] = \lim_{A\to\infty}  (1/A) \int_AV(\textbf{r})\exp(-i\textbf{q}\cdot\textbf{r})d\textbf{r}$ which is compatible with the one used in Ref. \cite{cao2019np} which only involves the periodic substrate.
We consider a cluster whose particles are located on a subset $\mathcal{L}_N$ of a two-dimensional lattice $\mathcal{L}$ with particle positions $\textbf{r}_j = j_1\textbf{a}_1 + j_2\textbf{a}_2 \stackrel{\text{def}}{=} \textbf{R}_j$.
The cluster covers an area $NA_0$, where $A_0$ is the area of the primitive cell of the lattice.
Given an arbitrary rigid translation $\textbf{r}_\textrm{c}$ of the cluster's positions the particle positions become $\textbf{r}_j = \textbf{R}_j + \textbf{r}_\textrm{c}$, with $\{\textbf{R}_j\} = \mathcal{L}_N \subset \mathcal{L}$.
We can write the per-particle interaction energy as
\begin{equation}
\begin{split}
 U(\textbf{r}_\textrm{c}, \theta_\textrm{o}) &= \frac 1N \sum_j V(\textbf{r}_j)							\\
	   &= \frac 1N \sum_j V(\textbf{R}_j + \textbf{r}_\textrm{c})						\\
	   &\cong \lim_{N \to \infty} \frac 1N \sum_{\textbf{R} \in \mathcal{L}_N}V(\textbf{R} + \textbf{r}_\textrm{c})				\\
   	   &\cong \lim_{N \to \infty} \sum_{\textbf{Q} \in \mathcal{Q}}  \frac1{NA_0} \int_{NA_0}\!\!\!\!\!\!\!V(\textbf{r} + \textbf{r}_\textrm{c}) \exp(-i\textbf{Q}\cdot\textbf{r}) d\textbf{r} 		\\
   	   &=\sum_{\textbf{Q} \in \mathcal{Q}} \tilde{V}(\textbf{Q}) \exp(i\textbf{Q}\cdot\textbf{r}_\textrm{c})					\\
	   &= \sum_{\textbf{Q} \in \mathcal{Q} \cap \bm{\Sigma}}  \tilde{V}(\textbf{Q}) \exp(i\textbf{Q}\cdot\textbf{r}_\textrm{c})
\end{split}    
\end{equation} 	   
Here $\mathcal{Q}$ is the reciprocal lattice of $\mathcal{L}$.
Note that at the $3^{\rm th}$ line, we have taken the limit $N \to \infty$ to exploit the Poisson summation at the $4^{\rm th}$ line:
$\lim_{N \to \infty} \sum_{\textbf{R} \in \mathcal{L}_N} V(\textbf{R}) = \lim_{N \to \infty} (1/A_0) \sum_{\textbf{Q}  \in \mathcal{Q}} \int_{NA_0} V(\textbf{r})\exp(-i\textbf{Q}\cdot\textbf{r})d\textbf{r}$.
The last passage stems from the fact that $\tilde{V}(\textbf{Q})$ is non-zero only if $\textbf{Q} \in \bm{\Sigma}$ is also satisfied, so that the only terms that contribute to the interaction energy are those in the intersection of the two reciprocal lattices. 

In addition to the observed directional locking of a crystalline cluster sliding across a quasiperiodic surface in Fig.~7(a), numerical simulations reported in Supplementary Fig.~S3 indicate that directional locking can also be observed in the reversed configuration, when a quasicrystalline cluster slides across a periodic surface.

\section{Conclusions}
In summary, we have experimentally observed and geometrically rationalized orientational and directional locking of close-packed colloidal crystalline clusters not just on triangular surfaces, but on more general cases, exemplified by square crystalline and by quasiperiodic substrates.
An elegant implicit relation exists between the locking orientation and the locked sliding direction, determined by the moir\'e pattern in real space and in reciprocal space respectively. The theoretical approach, which is purely geometrical, applies to rigid contacts with negligible  elastic deformations. 
The effect of elastic distortion, however, needs not to be only detrimental to the locking phenomena. Local distortions can affect the interaction between incommensurate crystals \cite{mctague1979prb,de2012prb,guerra2016nano}, which could help stabilize orientational and directional locking when the mismatches in CLVs are small.
The effect of finite size is an interesting topic of discussion in itself, to be addressed in a separate work.
This kind of orientational and directional locking effects will apply, barring complications due to size, shape, and edge effects, in nanomanipulation experiments where gold nanoparticles, flakes of graphene or other 2D materials, C60 islands, etc., are pushed across arbitrary crystal surfaces or quasicrystal substrates at low temperature.
More generally, they should also be relevant to the rheology of metacrystalline systems where crystalline monolayers with various periodic and quasiperiodic structures are built and investigated \cite{jiang2019nc,lee2020ac}.

\section{Acknowledgement}
X.C. acknowledges funding from Alexander von Humboldt Foundation. E.T. acknowledges support by ERC Advanced Grant ULTRADISS Contract No. 8344023. N.M. and A.V. acknowledge support by the Italian Ministry of University and Research through PRIN UTFROM N. 20178PZCB5.

\appendix

\section{
The relation between moir\'e patterns in real and reciprocal space for a triangular-lattice cluster on a square-lattice surface.}

To derive the relation between the real-space coincidence indexes $(m_1, m_2, n_1, n_2)$ and those in reciprocal space, $(M_1, M_2, N_1, N_2)$, for a triangular cluster on top of a square-lattice surface, we rewrite equation~(\ref{eq1}) as:
\begin{equation} \label{eqB1}
\begin{split}
    (\frac{4m_1}{3} + \frac{2m_2}{3})(\textbf{a}_1 - \frac{\textbf{a}_2}{2}) &\\
    + (\frac{2m_1}{3} + \frac{4m_2}{3})(\textbf{a}_2 - \frac{\textbf{a}_1}{2})&= n_1\textbf{b}_1 + n_2\textbf{b}_2.		
\end{split}    
\end{equation}

For a triangular lattice, $\textbf{a}_1 - \textbf{a}_2/2 = (\sqrt{3}a/2)\bm{\alpha}_1/|\bm{\alpha}_1|$, $\textbf{a}_2 - \textbf{a}_1/2 = (\sqrt{3}a/2)\bm{\alpha}_2/|\bm{\alpha}_2|$, $|\bm{\alpha}_1| = |\bm{\alpha}_2| = 2\pi / (\sqrt{3}a/2)$; for a square lattice, $\textbf{b}_1 = b\bm{\beta}_1 / |\bm{\beta}_1|$, $\textbf{b}_2 = b\bm{\beta}_2 / |\bm{\beta}_2|$, $|\bm{\beta}_1| = |\bm{\beta}_2| = 2\pi / b$. Inserting these relations into equation~(\ref{eqB1}) and considering that $b^2/a^2 = (m_1^2+m_2^2+m_1m_2) / (n_1^2+n_2^2)$, we obtain 
\begin{equation} \label{eqB2}
\begin{split}
  &(2m_1 + m_2)(n_1^2+n_2^2)\bm{\alpha}_1 + (m_1 + 2m_2)(n_1^2+n_2^2)\bm{\alpha}_2  \\
  &= (m_1^2+m_2^2+m_1m_2)(2n_1\bm{\beta}_1 + 2n_2\bm{\beta}_2). 
\end{split}    
\end{equation}
We see that equation~(\ref{eqB2}) provides the CLV in reciprocal space. To obtain the shortest reciprocal CLV, we divide a common integer factor $K$ to both sides of equation~(\ref{eqB2}), obtaining:
\begin{equation} \label{eqB3}
\begin{split}
      M_1 &= \frac{(2m_1+m_2)(n_1^2+n_2^2)}{K}, \\
      M_2 &= \frac{(m_1+2m_2)(n_1^2+n_2^2)}{K},	\\
      N_1 &= \frac{2n_1(m_1^2+m_2^2+m_1m_2)}{K},	\\
      N_2 &= \frac{2n_2(m_1^2+m_2^2+m_1m_2)}{K}.	       	  
\end{split}    
\end{equation}      
Here $K$ is an integer, such that $M_1$, $M_2$, $N_1$, $N_2$ are mutually relatively prime integers. Note that the vector Eq.(A2) is parallel to the vector Eq.(A1), therefore the CLVs in real space and in reciprocal space are parallel to one another for the triangle-on-square contact.

\section{
Orientational and directional locking with arbitrary two-dimensional lattices.}

We consider the situation of a two-dimensional lattice with primitive vectors $\textbf{a}_1$, $\textbf{a}_2$ sliding on top of another two-dimensional lattice with primitive vectors $\textbf{b}_1$, $\textbf{b}_2$. For arbitrary lattice spacings $|\textbf{a}_1|$, $|\textbf{a}_2|$, $|\textbf{b}_1|$, $|\textbf{b}_2|$ and arbitrary lattice angles $\phi = \arccos[(\textbf{a}_1\cdot\textbf{a}_2)/(|\textbf{a}_1||\textbf{a}_2|)]$, $\psi = \arccos[(\textbf{b}_1\cdot\textbf{b}_2)/(|\textbf{b}_1||\textbf{b}_2|)]$, given  a CLV in real-space moir\'e pattern:
\begin{equation} \label{eqC1}
m_1 \textbf{a}_1 + m_2 \textbf{a}_2 = n_1 \textbf{b}_1 + n_2 \textbf{b}_2.                   		   
\end{equation}  
we derive the necessary relations between the integers ($M_1$, $M_2$, $N_1$, $N_2$) and ($m_1$, $m_2$, $n_1$, $n_2$), such that $M_1 \bm{\alpha}_1 + M_2 \bm{\alpha}_2 = N_1 \bm{\beta}_1 + N_2 \bm{\beta}_2$ is a CLV in reciprocal space. As above, $\bm{\alpha}_1$, $\bm{\alpha}_2$, $\bm{\beta}_1$, $\bm{\beta}_2$ are the reciprocal primitive vectors that satisfy $\bm{\alpha}_i\cdot\textbf{a}_j=2\pi\delta_{ij}$ and $\bm{\beta}_i\cdot\textbf{b}_j=2\pi\delta_{ij}$ for $i,j=1,2$, and we make use of the angles $\phi$ and $\psi$ between $\textbf{a}_1$ and $\textbf{a}_2$ and between $\textbf{b}_1$ and $\textbf{b}_2$, respectively. For simplicity, we start our derivation with the equilateral case where $|\textbf{a}_2|/|\textbf{a}_1| = 1$ and $|\textbf{b}_2|/|\textbf{b}_1| = 1$. We then generalize the results to the case where $|\textbf{a}_2|/|\textbf{a}_1|$ and $|\textbf{b}_2|/|\textbf{b}_1|$ are not necessarily equal to 1. 

In the equilateral case, let $a = |\textbf{a}_1| = |\textbf{a}_2|$ and $b = |\textbf{b}_1| = |\textbf{b}_2|$, we can rewrite equation~(\ref{eqC1}) as:
\begin{equation} \label{eqC2}
\begin{split}
 &~(m_1 + m_2\cos\phi)(\textbf{a}_1 - \textbf{a}_2\cos\phi)\csc^2\phi \\
 &+ (m_2 + m_1\cos\phi)(\textbf{a}_2 - \textbf{a}_1\cos\phi)\csc^2\phi  \\
=&~(n_1 + n_2\cos\psi)(\textbf{b}_1 - \textbf{b}_2\cos\psi)\csc^2\psi \\
 &+ (n_2 + n_1\cos\psi)(\textbf{b}_2 - \textbf{b}_1\cos\psi)\csc^2\psi.       	    
\end{split}    
\end{equation} 
Note that 
\begin{equation} \label{eqC3}
\begin{split}
\textbf{a}_1 - \textbf{a}_2\cos\phi &= a\sin\phi\frac{\bm{\alpha}_1}{|\bm{\alpha}_1|}, \\
\textbf{a}_2 - \textbf{a}_1\cos\phi &= a\sin\phi\frac{\bm{\alpha}_2}{|\bm{\alpha}_2|}, \\
\textbf{b}_1 - \textbf{b}_2\cos\psi &= b\sin\psi\frac{\bm{\beta}_1}{|\bm{\beta}_1|}, \\
\textbf{b}_2 - \textbf{b}_1\cos\psi &= b\sin\psi\frac{\bm{\beta}_2}{|\bm{\beta}_2|}.	
\end{split}    
\end{equation} 
Replacing the corresponding terms in equation~(\ref{eqC2}) with the terms in equation~(\ref{eqC3}), we obtain:
\begin{equation} \label{eqC4}
\begin{split}
&(m_1 + m_2\cos\phi)\bm{\alpha}_1 + (m_2 + m_1\cos\phi)\bm{\alpha}_2 \\
=~&[(n_1 + n_2\cos\psi)\bm{\beta}_1 + (n_2 + n_1\cos\psi)\bm{\beta}_2]\frac{b^2}{a^2}.
\end{split}    
\end{equation} 
Here, the lattice spacing ratio must satisfy the condition $(b/a)^2 = (m_1^2 + m_2^2 + 2m_1m_2\cos\phi)/(n_1^2 + n_2^2 + 2n_1n_2\cos\psi)$ derived from equation~(\ref{eqC1}). 

Equation~(\ref{eqC4}) provides a CLV in reciprocal space with the same direction as that of equation~(\ref{eqC1}) if the corresponding four coefficients $(m_1 + m_2\cos\phi)$, $(m_2 + m_1\cos\phi)$, $(n_1 + n_2\cos\psi)b^2/a^2$, $(n_2 + n_1\cos\psi)b^2/a^2$ are all integers or can all be made integers by multiplying a suitable real common factor. This is immediately satisfied when $\phi$, $\psi$ = 60$^{\circ}$ or 90$^{\circ}$ (i.e. only triangular lattice or square lattice are involved), where $\cos\phi$, $\cos\psi$ = 1/2 or 0 are rational numbers. For a more general case (e.g. $\phi=45^{\circ}$ and $\psi = 90^{\circ}$), the corresponding coefficients in equation~(\ref{eqC4}) may not all become integers simultaneously. We therefore apply a rotational operator $\mathcal{R}_{90}$ to equation~(\ref{eqC4}) to obtain a perpendicular component, which is:
\begin{equation} \label{eqC5}
\begin{split}
 &(m_1 + m_2\cos\phi)\mathcal{R}_{90}\bm{\alpha}_1 + (m_2 + m_1\cos\phi)\mathcal{R}_{90}\bm{\alpha}_2 \\
=&~[(n_1 + n_2\cos\psi)\mathcal{R}_{90}\bm{\beta}_1 + (n_2 + n_1\cos\psi)\mathcal{R}_{90}\bm{\beta}_2]\frac{b^2}{a^2}.
\end{split}    
\end{equation} 
Here $\mathcal{R}_{90}$ rotates all the vectors by 90$^{\circ}$ in the anti clockwise direction, i.e.
\begin{equation} \label{eqC6}
\begin{split}
\mathcal{R}_{90}\bm{\alpha}_1 &= (\bm{\alpha}_2 + \bm{\alpha}_1\cos\phi)\csc\phi, \\
\mathcal{R}_{90}\bm{\alpha}_2 &= -(\bm{\alpha}_1 + \bm{\alpha}_2\cos\phi)\csc\phi, \\
\mathcal{R}_{90}\bm{\beta}_1  &= (\bm{\beta}_2 + \bm{\beta}_1\cos\psi)\csc\psi, \\
\mathcal{R}_{90}\bm{\beta}_2  &= -(\bm{\beta}_1 + \bm{\beta}_2\cos\psi)\csc\psi. 
\end{split}    
\end{equation} 
Replacing the corresponding terms in equation~(\ref{eqC5}) with the terms in equation~(\ref{eqC6}) we obtain:
\begin{equation} \label{eqC7}
\begin{split}
 &-m_2\sin\phi\bm{\alpha}_1 + m_1\sin\phi\bm{\alpha}_2 \\
 =&~ (-n_2\sin\psi\bm{\beta}_1 + n_1\sin\psi\bm{\beta}_2)\frac{b^2}{a^2}.
\end{split}    
\end{equation} 

Similar to equation~(\ref{eqC4}), equation~(\ref{eqC7}) is a CLV in reciprocal space in the perpendicular direction (as a result of the operator $\mathcal{R}_{90}$) of equation~(\ref{eqC1}), if the corresponding coefficients  $-m_2\sin\phi$, $m_1\sin\phi$, $-n_2\sin\psi b^2/a^2$, $n_1\sin\psi b^2/a^2$ are all integer numbers or can all become integer numbers by multiplying a suitable common factor.
This is also not necessarily true.
However, now we can construct a linear combination $\kappa$(\ref{eqC4}) + $\gamma$(\ref{eqC7}), such that, by choosing proper $\kappa$ and $\gamma$, the new equation 
\begin{equation} \label{eqC8}
\begin{split}
&[\kappa(m_1 + m_2\cos\phi) - \gamma m_2\sin\phi]\bm{\alpha}_1  \\
&+ [\kappa(m_2 + m_1\cos\phi) + \gamma m_1\sin\phi]\bm{\alpha}_2 \\ 
= & [\kappa(n_1 + n_2\cos\psi) - \gamma n_2\sin\psi]\frac{b^2}{a^2}\bm{\beta}_1 \\
&+ [\kappa(n_2 + n_1\cos\psi) + \gamma n_1\sin\psi]\frac{b^2}{a^2}\bm{\beta}_2.
\end{split}    
\end{equation} 
could provide a CLV in reciprocal space. This requires that the coefficients $\kappa(m_1 + m_2\cos\phi) - \gamma m_2\sin\phi$, $\kappa(m_2 + m_1\cos\phi) + \gamma m_1\sin\phi$, $[\kappa(n_1 + n_2\cos\psi) - \gamma n_2\sin\psi]b^2/a^2$ and $[\kappa(n_2 + n_1\cos\psi) + \gamma n_1\sin\psi]b^2/a^2$ are all integers.
Considering that $b^2/a^2 = (m_1^2 + m_2^2 + 2m_1m_2\cos\phi)/(n_1^2 + n_2^2 + 2n_1n_2\cos\psi)$, we could let
\begin{equation} \label{eqC9}
\begin{split}
&\kappa(m_1 + m_2\cos\phi) - \gamma m_2\sin\phi \\
=&~ M_1(m_1^2 + m_2^2 + 2m_1m_2\cos\phi), \\
&\kappa(m_2 + m_1\cos\phi) + \gamma m_1\sin\phi \\
=&~ M_2(m_1^2 + m_2^2 + 2m_1m_2\cos\phi), \\
&\kappa(n_1 + n_2\cos\psi) - \gamma n_2\sin\psi \\
=&~ N_1(n_1^2 + n_2^2 + 2n_1n_2\cos\psi), \\
&\kappa(n_2 + n_1\cos\psi) + \gamma n_1\sin\psi \\
=&~ N_2(n_1^2 + n_2^2 + 2n_1n_2\cos\psi).
\end{split}    
\end{equation} 
In this way
equation~(\ref{eqC8}) becomes
\begin{equation} \label{eqC10}
M_1\bm{\alpha}_1 + M_2\bm{\alpha}_2 = N_1\bm{\beta}_1 + N_2\bm{\beta}_2.  		
\end{equation} 		    
This requires
\begin{equation} \label{eqC11}
\kappa = M_1m_1 + M_2m_2 = N_1n_1 + N_2n_2.
\end{equation} 	
and 
\begin{equation} \label{eqC12}
\begin{split}
\gamma &= [M_2(m_1 + m_2\cos\phi) - M_1(m_2 + m_1\cos\phi)]\csc\phi \\
&= [N_2(n_1 + n_2\cos\psi) - N_1(n_2 + n_1\cos\psi)]\csc\psi.
\end{split}    
\end{equation} 
Let 
\begin{equation} \label{eqC13}
\begin{split}
M_1 &= M_{11}n_1 + M_{12}n_2, \\
M_2 &= M_{21}n_1 + M_{22}n_2, \\
N_1 &= N_{11}m_1 + N_{12}m_2, \\
N_2 &= N_{21}m_1 + N_{22}m_2.				
\end{split}    
\end{equation} 
Inserting equation~(\ref{eqC13}) into equation~(\ref{eqC11}) we have:
\begin{equation} \label{eqC14}
\begin{split}
&M_{11}m_1n_1 + M_{12}m_1n_2 + M_{21}m_2n_1 + M_{22}m_2n_2 \\
=&~ N_{11}n_1m_1 + N_{12}n_1m_2 + N_{21}n_2m_1 + N_{22}n_2m_2.
\end{split}    
\end{equation} 
We see that, a simple choice for us is 
\begin{equation} \label{eqC15}
\begin{split}
M_{11} &= N_{11}, \\
M_{12} &= N_{21}, \\
M_{21} &= N_{12}, \\
M_{22} &= N_{22}. 	
\end{split}    
\end{equation} 			  
Similarly, inserting equation~(\ref{eqC13}) into equation~(\ref{eqC12}) allows us to obtain
\begin{equation} \label{eqC16}
\begin{split}
(M_{21} - M_{11}\cos\phi)\csc\phi &= (N_{21} - N_{11}\cos\psi)\csc\psi, \\
(M_{22} - M_{12}\cos\phi)\csc\phi &= (N_{21}\cos\psi - N_{11})\csc\psi, \\
(M_{21}\cos\phi - M_{11})\csc\phi &= (N_{22} - N_{12}\cos\psi)\csc\psi, \\
(M_{22}\cos\phi - M_{12})\csc\phi &= (N_{22}\cos\psi - N_{12})\csc\psi.
\end{split}    
\end{equation} 

Equation~(\ref{eqC15}) and (\ref{eqC16}) are a total of 8 equations. Only 6 of them are independent. 
Let $M_{11} = \epsilon$, $M_{12} = \eta$. Then
\begin{equation} \label{eqC17}
\begin{split}
	M_{21} &= [\eta\sin\phi - \epsilon\sin(\phi-\psi)]\csc\psi, \\
	M_{22} &= [-\epsilon\sin\phi + \eta\sin(\phi+\psi)]\csc\psi.
\end{split}    
\end{equation} 
$N_{11}$, $N_{21}$, $N_{12}$, $N_{22}$ can be obtained from equation~(\ref{eqC15}).

Inserting the 8 coefficients $M_{11}$, $M_{12}$, $M_{21}$, $M_{21}$, $N_{11}$, $N_{12}$, $N_{21}$, $N_{21}$ into equation~(\ref{eqC13}), we obtain	
\begin{equation} \label{eqC18}
\begin{split}
M_1 &= \epsilon n_1 + \eta n_2, \\
M_2 &= \frac{\eta \sin\phi - \epsilon\sin(\phi-\psi)}{\sin\psi} n_1 \\
	&+ \frac{-\epsilon\sin\phi + \eta\sin(\phi+\psi)}{\sin\psi} n_2, \\
N_1 &= \epsilon m_1 + \frac{\eta\sin\phi - \epsilon\sin(\phi-\psi)}{\sin\psi} m_2, \\
N_2 &= \eta m_1 + \frac{-\epsilon\sin\phi + \eta\sin(\phi+\psi)}{\sin\psi} m_2.
\end{split}    
\end{equation} 
or
\begin{equation} \label{eqC19}
\begin{split}
&(\epsilon n_1 + \eta n_2)\bm{\alpha}_1 + [\frac{\eta \sin\phi - \epsilon\sin(\phi-\psi)}{\sin\psi} n_1 \\
&+ \frac{-\epsilon\sin\phi + \eta\sin(\phi+\psi)}{\sin\psi} n_2]\bm{\alpha}_2 \\
&= [\epsilon m_1 + \frac{\eta\sin\phi - \epsilon\sin(\phi-\psi)}{\sin\psi} m_2]\bm{\beta}_1 \\
&+ [\eta m_1 + \frac{-\epsilon\sin\phi + \eta\sin(\phi+\psi)}{\sin\psi} m_2]\bm{\beta}_2. 
\end{split}    
\end{equation} 

Equation~(\ref{eqC19}) is the CLV in reciprocal space if we could find proper real numbers $\epsilon$ and $\eta$ such that all the coefficients in equation~(\ref{eqC19}) are integers. 

As an example, for two lattices of the same mutual angle $\phi=\psi$, but different spacings, equation~(\ref{eqC19}) becomes:
\begin{equation} \label{eqC20}
\begin{split}
&(\epsilon n_1 + \eta n_2)\bm{\alpha}_1 + (\eta n_1 - \epsilon n_2 + 2\eta n_2\cos\phi)\bm{\alpha}_2 \\
&= (\epsilon m_1 + \eta m_2)\bm{\beta}_1 + (\eta m_1 - \epsilon m_2 + 2\eta m_2\cos\phi)\bm{\beta}_2.
\end{split}    
\end{equation} 
If we choose $\eta=0$, $\epsilon=1$, we obtain:
\begin{equation} \label{eqC21}
n_1\bm{\alpha}_1 - n_2\bm{\alpha}_2 = m_1\bm{\beta}_1 - m_2\bm{\beta}_2. 
\end{equation} 
This is the formula we obtained in our previous publication \cite{cao2019np} for a triangular cluster on another triangular cluster. Here we see that it also applies for non-triangular equilateral lattices as long as the primitive vectors in two lattices form the same angle.

As another example, for two lattices with $\phi=45^{\circ}$ and $\psi=90^{\circ}$, and a CLV $\textbf{a}_1 + 2\textbf{a}_2 = \textbf{b}_1 + \textbf{b}_2$ in real space. Inserting $m_1=1$, $m_2=2$, $n_1=1$, $n_2=1$, $\phi=45^{\circ}$ and $\psi=90^{\circ}$ into equation~(\ref{eqC19}) we obtain
\begin{equation} \label{eqC22}
(\epsilon + \eta)\bm{\alpha}_1 + \sqrt{2}\eta\bm{\alpha}_2 = [\epsilon+\sqrt{2}(\epsilon + \eta)]\bm{\beta}_1 + [\eta+\sqrt{2}(\eta - \epsilon)]\bm{\beta}_2.
\end{equation} 
Now, if we choose $\epsilon=1-\sqrt{2}$ and $\eta = \sqrt{2}$, we obtain
\begin{equation} \label{eqC23}
\bm{\alpha}_1 + 2\bm{\alpha}_2 = \bm{\beta}_1 + 4\bm{\beta}_2. 
\end{equation} 
Note that the CLVs in reciprocal space $\bm{\alpha}_1 + 2\bm{\alpha}_2 = \bm{\beta}_1 + 4\bm{\beta}_2$ and in real space $\textbf{a}_1 + 2\textbf{a}_2 = \textbf{b}_1 + \textbf{b}_2$ are neither parallel nor perpendicular to one another. 

Finally, to generalize equation~(\ref{eqC18}) and (\ref{eqC19}) to arbitrary 2D lattices with $\lambda_\textrm{a} = |\textbf{a}_2|/|\textbf{a}_1|$ 
and $\lambda_\textrm{b} = |\textbf{b}_2|/|\textbf{b}_1|$,
we will need to substitute all instances of $m_2$, $M_2$, $\bm{\alpha}_2$, $n_2$, $N_2$, and $\bm{\beta}_2$ in the relevant equations with $\lambda_\textrm{a}m_2$, $\lambda_\textrm{a}^{-1}M_2$, $\lambda_\textrm{a}\bm{\alpha}_2$, $\lambda_\textrm{b}n_2$, $\lambda_\textrm{b}^{-1}N_2$, and $\lambda_\textrm{b}\bm{\beta}_2$ respectively. The result is:

\begin{equation} \label{eqC24}
\begin{split}
      M_1 &= \epsilon n_1 + \eta \lambda_\textrm{b} n_2, \\
      M_2 &= [\frac{\eta\sin\phi-\epsilon\sin(\phi-\psi)}{\sin\psi}n_1 \\
      	  &+ \frac{-\epsilon\sin\phi + \eta\sin(\phi+\psi)}{\sin\psi}\lambda_\textrm{b}n_2]\lambda_\textrm{a},	\\
      N_1 &= \epsilon m_1 + \frac{\eta\sin\phi-\epsilon\sin(\phi-\psi)}{\sin\psi}\lambda_\textrm{a} m_2,	\\
      N_2 &= [\eta m_1 + \frac{-\epsilon\sin\phi + \eta\sin(\phi+\psi)}{\sin\psi}\lambda_\textrm{a}m_2]\lambda_\textrm{b}.
\end{split}
\end{equation} 
and

\begin{equation} \label{eqC25}
\begin{split}
&(\epsilon n_1 + \eta\lambda_\textrm{b} n_2)\bm{\alpha}_1 + [\frac{\eta \sin\phi - \epsilon\sin(\phi-\psi)}{\sin\psi}n_1 \\
&+ \frac{-\epsilon\sin\phi + \eta\sin(\phi+\psi)}{\sin\psi}\lambda_\textrm{b}n_2]\lambda_\textrm{a}\bm{\alpha}_2 \\
&= [\epsilon m_1 + \frac{\eta\sin\phi - \epsilon\sin(\phi-\psi)}{\sin\psi}\lambda_\textrm{a}m_2]\bm{\beta}_1 \\
&+ [\eta m_1 + \frac{-\epsilon\sin\phi + \eta\sin(\phi+\psi)}{\sin\psi}\lambda_\textrm{a}m_2]\lambda_\textrm{b}\bm{\beta}_2. 
\end{split}    
\end{equation} 
Note that, for arbitrary lattices, there might not be integer solutions for $(M_1,M_2,N_1,N_2)$. Nevertheless, we can always find proper $\epsilon$ and $\eta$ such that $(M_1,M_2,N_1,N_2)$ are all approximately integers, as we have done for the case illustrated in Fig.~6.

\bibliographystyle{apsrev}

\end{document}

\title{Supplementary Information for "Pervasive orientational and directional locking at geometrically heterogeneous sliding interfaces"}
\author{Xin Cao$^{1}$}
\author{Emanuele Panizon$^{1}$}
\author{Andrea Vanossi$^{2,3}$}
\author{Nicola Manini$^{4}$}
\author{Erio Tosatti$^{2,3,5}$}
\author{Clemens Bechinger$^{1}$}
\email{clemens.bechinger@uni-konstanz.de}
\affiliation{$^1$Fachbereich Physik, University Konstanz, 78464 Konstanz, Germany}
\affiliation{$^2$International School for Advanced Studies (SISSA), Via Bonomea 265, 34136 Trieste, Italy}
\affiliation{$^3$CNR-IOM Democritos National Simulation Center, Via Bonomea 265, 34136 Trieste, Italy}
\affiliation{$^4$Dipartimento di Fisica, Universita degli Studi di Milano, Via Celoria 16, 20133 Milano, Italy}
\affiliation{$^5$International Centre for Theoretical Physics (ICTP), Strada Costiera 11, 34151 Trieste, Italy}
\date{\today}

\maketitle

\makeatletter
\renewcommand{\thefigure}{S\@arabic\c@figure}
\makeatother
\setcounter{figure}{0}

\begin{figure}
  \centering
  \includegraphics[width=0.8\columnwidth]{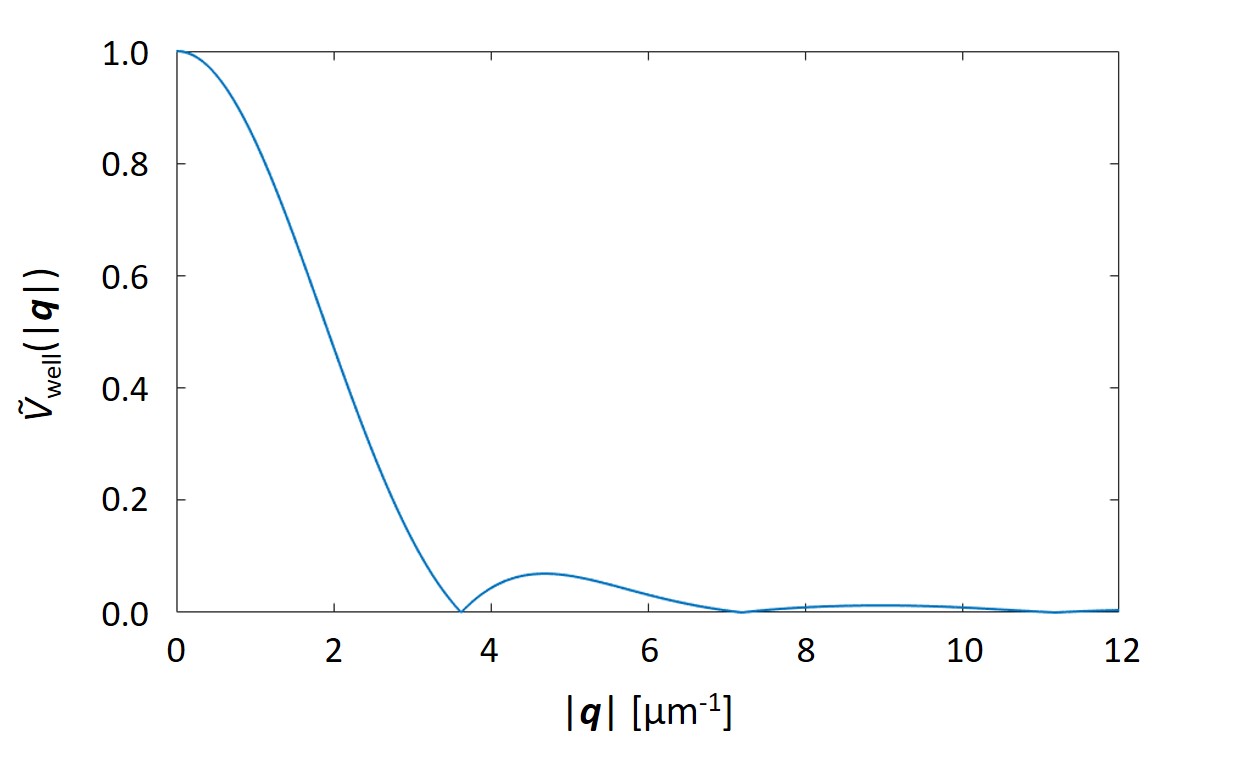}
\caption{The Fourier transform $\tilde{V}_\textrm{well}(|\textbf{q}|)$ of the single-well potential $V_\textrm{well}(|\textbf{r}|)$ expressed in the main text of our manuscript. The strength of the Fourier transform decays to $\sim$0.4 of its maximum value at $|\textbf{q}| = 2 ~\mu\textrm{m}^{-1}$.}
\label{figs1}
\end{figure}

\begin{figure}
  \centering
  \includegraphics[width=1.0\columnwidth]{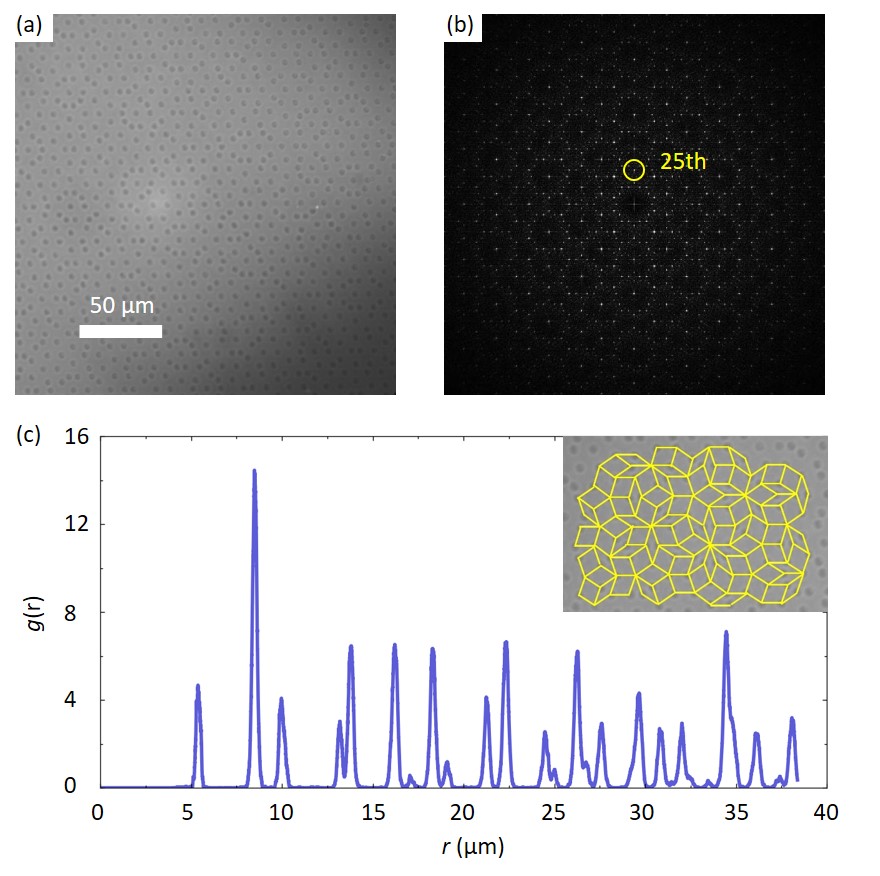}
\caption{\textbf{Structure of the quasiperiodic surface.} (a) A microscope image of the quasiperiodically corrugated surface. (b) The center region of the Fourier transform of the quasiperiodic pattern in (a), showing the 10-fold rotational symmetry. The $25^{th}$ peak is highlighted. The image reveals the peak intensity in log scale. A black background is used so that to have a better contrast of the scattering peaks. (c) The radial distribution function $g(r)$ for the quasiperiodic surface, showing a crystal-like structure. The ratio between the positions of the second peak and of the first peak is the golden ratio. (Inset) The rhombus tiling structure of the quasiperiodic surface. The lateral length of the rhombus is 8.51 $\mu\textrm{m}$. The quasiperiodic lattice is generated with a dual grid method described in references \cite{su2017jcp,socolar1985prb}.}
\label{figs2}
\end{figure}

\begin{figure}
  \centering
  \includegraphics[width=1.0\columnwidth]{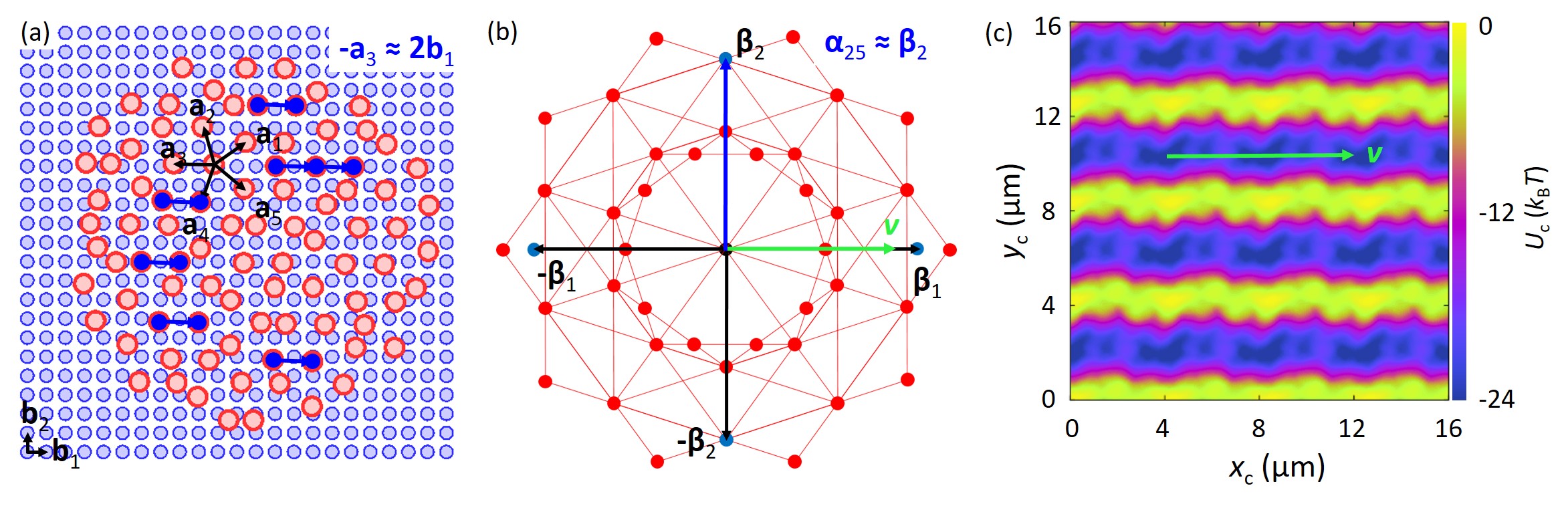}
\caption{\textbf{Orientational and directional locking of a quasicrystalline cluster on a square surface.} (a) A quasicrystalline cluster (red) on top of a square-lattice surface (blue). The quasicrystalline cluster has the same structure as the quasiperiodic surface in Fig. 7(a). The generating vectors $|\textbf{a}_1| = |\textbf{a}_2| = |\textbf{a}_3| = |\textbf{a}_4| = |\textbf{a}_5| = a = 8.51 ~\mu\textrm{m}$ and form $72^{\circ}$ mutual angles; for the substrate $|\textbf{b}_1| = |\textbf{b}_2| =  b = 4.16 ~\mu\textrm{m}$. The 'moir\'e pattern' is characterised by the smallest CLV $\textrm{-}\textbf{a}_3 = 2\textbf{b}_1$ indicated by the blue line. (b) The 'moir\'e pattern' of the corresponding reciprocal lattices in (a). The blue line indicates  $\bm{\alpha}_{25} = \bm{\beta}_2$, where $\bm{\alpha}_{25}$ is the 25$^\textrm{th}$ peak in the Fourier transform of the quasiperiodic structure highlighted in Fig. S2(b). (c) The calculated per-particle potential as a function of the center-of-mass position $x_\textrm{c}$ and $y_\textrm{c}$ of the quasicrystalline cluster in (a). The potential energy landscape indicates (green arrow) a low energy corridor along $0^{\circ}$ direction, perpendicular to the reciprocal CLV in (b), as expected.}
\label{figs3}
\end{figure}

\href{https://www.dropbox.com/s/2935zpp7zjnpl1t/movie1.avi?dl=0}{\textbf{Movie 1.}} Orientational and directional locking observed experimentally for the colloidal cluster in Fig. 1(b) of the main text on top of a $b = 5.0~\mu\textrm{m}$ square-lattice patterned surface. The driving force $F$ = 72 fN from left to right horizontally. Movie acceleration: 40 $\times$ real time.

\href{https://www.dropbox.com/s/akipi4h4pkd4ico/movie2.avi?dl=0}{\textbf{Movie 2.}} Orientational and directional locking observed experimentally for colloidal clusters with various sizes and shapes on top of a $b = 5.0~\mu\textrm{m}$ square-lattice patterned surface, as in Fig. 1 of the main text. The driving force $F$ = 72 fN from left to right horizontally. Movie acceleration: 40 $\times$ real time.

\href{https://www.dropbox.com/s/gp0adfl750zwlsg/movie3.avi?dl=0}{\textbf{Movie 3.}} Orientational and directional locking observed experimentally for a colloidal cluster on top of a $b = 4.8~\mu\textrm{m}$ square-lattice patterned surface, as in Fig. 3 of the main text. The driving force $F$ = 89 fN from left to right horizontally. Movie acceleration: 40 $\times$ real time.

\href{https://www.dropbox.com/s/hehxh3otbc7uxmi/movie4.avi?dl=0}{\textbf{Movie 4.}} Orientational and directional locking observed experimentally for a colloidal cluster on top of a $b = 5.4~\mu\textrm{m}$ square-lattice patterned surface, as in Fig. 4 of the main text. The driving force $F$ = 89 fN from left to right horizontally. Movie acceleration: 40 $\times$ real time.

\href{https://www.dropbox.com/s/wiwy5hihl9x6oa7/movie5.avi?dl=0}{\textbf{Movie 5.}} Orientational and directional locking observed experimentally for a colloidal cluster on top of a $b = 6.2~\mu\textrm{m}$ square-lattice patterned surface, as in Fig. 5 of the main text. The driving force $F$ = 89 fN from left to right horizontally. Movie acceleration: 40 $\times$ real time.

\href{https://www.dropbox.com/s/ajs96fu6ficltit/movie6.avi?dl=0}{\textbf{Movie 6.}} Orientational and directional locking observed experimentally for the colloidal cluster in Fig. 7(a) of the main text on top of a quasiperiodically patterned surface. The driving force $F$ = 72 fN from left to right horizontally. Movie acceleration: 40 $\times$ real time.

\bibliographystyle{apsrev}